\documentclass[12pt]{article}
\usepackage{graphicx}
\let\ps=\psi

\def\\{\hfill\break} \let\==\equiv

\let\0=\noindent
\def\ket#1{{|#1\rangle}}

\def\nn{\nonumber}

\def\qed{\hfill\raise1pt\hbox{\vrule height5pt width5pt depth0pt}}

\def\be{\begin{equation}}
\def\ee{\end{equation}}
\def\bea{\begin{eqnarray}}\def\eea{\end{eqnarray}}
\def\be{\begin{equation}}
\def\ee{\end{equation}}
\def\bea{\begin{eqnarray}}
\def\eea{\end{eqnarray}}

\def\hsp5{\hspace{5mm}}

\def\case#1/#2{\textstyle\frac{#1}{#2}}
\title{The arrow of time and the nature of spacetime}
\author{George F R Ellis,\\
Mathematics Department, University of Cape Town.}

\begin{document}

\maketitle \vspace{0.1in}
\begin{abstract}
This paper extends the work of a previous paper \cite{Ell13} on the
flow of time, to consider the origin of the arrow of time. It
proposes that a `past condition' cascades down from cosmological to
micro scales, being realized in many microstructures and setting the
arrow of time at the quantum level by top-down causation. This
physics arrow of time then propagates up, through underlying
emergence of higher level structures, to geology, astronomy,
engineering, and biology. The appropriate space-time picture to view
all this is an emergent block universe (`EBU'), that recognizes the
way the present is different from both the past and the future. This
essential difference is the ultimate reason the arrow of time has to
be the way it is.
\end{abstract}
\tableofcontents

\section{Quantum theory and the arrow of time}
The arrow of time problem is one of the major foundational problems
in physics
\cite{WheFey45,EllSci72,Dav74,HalPerZur96,Pri96,Zeh07,Car10},
because the one-way flow of time embodied in the second law of
thermodynamics emerges from time-symmetric microphysics. This paper
builds on a previous paper \cite{Ell11a}, where the emergence of
higher level structures and their top-down influence on lower level
structures was considered to be a key factor in looking at the
quantum measurement problem and the nature of the classical quantum cut.
This paper will propose the same is true of the arrow of time.\\

The paper is structured as follows: Section \ref{sec:foundations}
sets up the basic ideas underlying the rest of the paper: the basics
of quantum theory, and the ideas of bottom up and top down causation
in the hierarchy of complexity, and makes a main proposal as to how
quantum theory underlies complex systems (Section \ref{prop}).
Section \ref{time2} discusses the arrow of time problem and various
proposals that have been made to solve it, focussing on the Past
Hypothesis in  Section \ref{time9}. Section \ref{time3} looks at the
cosmological context and the various epochs relevant to discussing
the issue of  the flow of time, and how the basic cosmological direction
of time is set up.  Section \ref{time4}  proposes that the
cosmological arrow of time cascades down to lower levels in the
hierarchy, each level in turn communicating the arrow to the next
lower level.  Section \ref{time6} discusses how the arrow of time
then flows up the hierarchy as emergent structures form out of lower
level elements,  indeed being taken for granted in this context.
Section \ref{sec:time1} considers how the natural spacetime view to
accommodate the experienced ongoing flow of time is an Emergent
Block Universe that grows with time, and where the past, present and
future are represented as having a quite different ontological
character from each other. This provides the ultimate rationale for
the arrow of time  (Section \ref{sec:arrow}) and resolves potential
problems arising from the possibility of closed timelike lines
(Section \ref{sec:closed}).  Thus the overall picture that emerges
is one of the arrow of time in physical systems being determined in
a top-down manner, starting off from a special initial condition at
the cosmological scale where the cosmological arrow time sets the
basic direction of causation, but then emerging in complex systems
through bottom up causation (Section \ref{time10}).  The relation to
state vector reduction is crucial; obviously the details of that
relation are still to be resolved (Section \ref{sec:conclusion}).

\section{Foundations}\label{sec:foundations}
To set the scene, this sections summarizes some foundational issues
discussed in more depth in \cite{Ell11a}. It considers basics of
quantum dynamics (Section \ref{sec:basic}), the hierarchy of
structure (Section \ref{sec:context}), and interlevel relations in
that structure  (Section \ref{relations}). It then puts the main
viewpoint underlying this and the companion paper (Section
\ref{main1}), emphasizing the interaction between bottom-up and
top-down causation as a key feature of physics,

\subsection{Quantum dynamics}\label{sec:basic}
The basic
postulate of quantum mechanics \cite{Mor90,Rae94,Ish95,GreZaj06} is that
a system is described by a state vector $\ket\ps$ that generically can be written as a linear
combination of unit orthogonal basis vectors
 \be
 \ket{\ps_1} = \sum_n c_n\ket{u_n(x)}, \label{wave}
 \ee
where $u_n$ are the eigenstates of some observable $\hat{A}$ (\cite{Ish95}:5-7).
The evolution of the system can be 
described by a unitary operator $\widehat{U}(t_2,t_1)$, and so
evolves as

\be \ket{\ps_2} = \widehat{U}(t_2,t_1)\,\ket{\ps_1} \label{U1} \ee

\noindent Here $\widehat{U}(t_2,t_1)$ is the standard
evolution operator, determined by the evolution equation

\be i\hbar\frac{d}{dt} \ket{\ps_t} = \hat{H}  \ket{\ps_t}.
\label{evolution} \ee When the Hamiltonian $\hat{H}$ is time
independent, $\widehat{U}$ has the form (\cite{Ish95}:102-103) \be
\widehat{U}(t_2,t_1) = e^{-\frac{i}{\hbar} \hat{H}(t_2-t_1)}
\label{U2}. \ee
As well as the unitary evolution (\ref{U1}), measurements take place. Immediately after a measurement is made at a time $t=t^*$,  the relevant part of the wavefunction is found to be in one of the
eigenstates:
 \be
 \ket{\ps_2} = c_N \ket{u_N(x)} \label{collapse}
 \ee
for some specific index $N$. The data for $t< t^*$ do not determine
either $N$ or $c_N$; they only determine a probability for each
possible outcome (\ref{collapse}) 
through the
fundamental equation
 \be
p_N = c_N^2 = \langle e_N | \psi_1\rangle^2.  \label{prob}
 \ee
One can think of this as due to the probabilistic time-irreversible
reduction of the wave function
\be
 \begin{array}{c c c}
 \ket{\ps_1} = \sum_n c_n\ket{u_n(x)} \hspace{1.75 cm}
 \longrightarrow \hspace{1.5 cm}\ket{\ps_2} = c_N u_N(x)\nn\\
 Indeterminate  \hspace{2 cm} Transition  \hspace{1 cm} Determinate \\  \label{trans}
\end{array}
\ee This is the event where the uncertainties of quantum theory
become manifest (up to this time the evolution is determinate and
time reversible). It will not be a unitary transformation (\ref{U1})
unless the initial state was already an eigenstate of $\hat{A}$.
More generally, one has projection into a subspace of eigenvectors
(\cite{Ish95}:136; \cite{WisMil10}:10-12) or a transformation of
density matrices (\cite{Ish95}:137), or any other of a large set of
possibilities (\cite{WisMil10}:8-42), but the essential feature of
non-unitary evolution remains the core of the process.\\

The process (\ref{trans}) is where the time irreversibility, and
hence the arrow of time, is manifested at the quantum level:
the eigenstate (\ref{collapse}) occurs at a later time than the
superposition (\ref{wave}), and knowledge of the final state (\ref{collapse}) does not determine the initial state (\ref{wave}); the values of the coefficients $u_n$ have been lost. After collapse the dynamics will tend to cause new superpositions (\ref{wave}) to emerge through the unitary process
(\ref{U1}); then further effective non-unitary wave vector reduction events will
produce eigenstates again. \\

There are other understandings of what happens when a measurement tales place, such as the
Everett many worlds theory \cite{Wal01}; however they have to lead
to an effective behavior as outline above, or they do not
correspond to experiment.

\subsection{The context: the hierarchy of the
structure}\label{sec:context} The context in which this all occurs
is the hierarchy of structure and causation
\cite{Ell08,Ell11,Ell11a}. Table 1 gives a simplified representation
of this hierarchy of levels of reality as characterized by
corresponding academic subjects, with the natural sciences on the
left and the life sciences on the right. On both sides, each lower
level underlies what happens at each higher level in terms of structure and causation.

\begin{center}
\begin{tabular}{|l|l|l|}\hline
 Level 10: &  Cosmology &  Sociology/Economics/Politics\\ \hline
  Level 9: &  Astronomy  &  Psychology \\ \hline
  Level 8: &  Space science &  Physiology\\ \hline
  Level 7: &  Geology, Earth science &  Cell biology \\ \hline
  Level 6: &  Materials science &  Biochemistry \\ \hline
  Level 5: &  Macro physics, physical chemistry &  Organic Chemistry \\ \hline
  Level 4: &  Atomic Physics &  Atomic Physics \\ \hline
  Level 3: &  Nuclear Physics &  Nuclear Physics \\ \hline
  Level 2: &  Particle physics &  Particle physics\\ \hline
  Level 1: &  Fundamental Theory &  Fundamental Theory  \\
  \hline
\end{tabular}\\
\end{center}

{}\\ \textbf{Table 1:} \emph{The hierarchy of structure and
causation for inanimate matter (left) and for life (right)}. For a
more detailed description of this hierarchical structure,
 see \\http://www.mth.uct.ac.za/$\sim$ellis/cos0.html.\\

The ordering of the levels on the left is by scale, which is also
the inverse of energy. However it also represents a putative
layering of emergent causation: for example one can claim that
particle physics underlies nuclear physics in that nuclei are made
of combinations of quarks;  nuclear physics underlies atomic
physics, in that atomic properties depend on nuclear properties;
atoms underly molecules which underlie the kinetic theory of gases;
and so on; each level emerges in this way from combinations of lower
level entities. In many cases the relevant higher level variables
are coarse-grained lower level variables  \cite{Ell11a}.\\

In the case of the life sciences  \cite{CamRee05}, ordering is by
physical scale (biochemistry underlies microbiology, which underlies
cell biology for example) and timescale (interactions are much
faster at lower levels) until the higher levels, where the nature of
causation changes because of the emergence of mind. Here it is
commonly thought that psychology emerges from physiology
\cite{Nicateal01}, and society from the interaction of individual minds. The
relationships between the different levels are very
different from one another in these cases; nevertheless the
hierarchy as presented makes sense as a hierarchy of emergent causal
levels, each with its own relevant variables and effective laws of behavior
\cite{And72}.

\subsection{Inter level relations}\label{relations}  It is useful to characterize causation in this hierarchical
context as proceeding in both a bottom-up and a top-down manner
\cite{Ell08,EllNobOCo11}.

\subsubsection{Bottom-up Effects}
Higher level structure emerges from combination of lower level
structural elements, for example molecules emerge from atoms
\cite{FeyLeiSan63}, with higher level dynamics emerging from lower
level dynamics through the effects of the lower level dynamics in
the context of the higher level emergent structure, for example
molecular biology emerges from physics \cite{Wat70}. Thus behavior
on level $X + 1$ emerges from behavior on level $X$.
Often there is \emph{coarse-graining} of lower level variables (e.g.
particle states) to give higher level variables (e.g. density and
pressure) and effective emergent laws (e.g. the perfect gas laws)
\cite{AloFin71}, accompanied by a conversion of usable to non-usable
energy when some energy is hidden in lower level states, and hence
not manipulable via higher level variables.

\subsubsection{The emergence of higher level behavior}\label{emergeh}
Consequent on bottom up causation, higher level behavior emerges
from that at the lower levels. Consider how higher level behavior
relates to lower level behavior in two adjacent levels in the
hierarchy of complexity (Diagram 1). \\

\begin{center}
\begin{tabular}{|l|c|l|c|}\hline
   Level $N+1$: & Initial state $I$ &  \emph{Higher level theory} $T$: $\Rightarrow$ & Final state $F$ \\ \hline
    & $\Uparrow $ &  \emph{\textbf{Coarse grain}} & $\Uparrow $ \\ \hline
  Level $N$:  & Initial state $i$ &  \emph{Lower level theory} $t$: $\Rightarrow$ &  Final state $f$ \\ \hline
\end{tabular}\\
\end{center}

{}\\ \textbf{Diagram 1:} \emph{The emergence of higher level
behavior from lower level theory. Coarse-graining the action of the
lower-level theory results in an effective higher level theory}.\\

The dynamics of the lower level theory maps an initial state $i$ to
a final state $f$. Coarse graining the lower level variables, state
$i$ corresponds to the higher level state $I$ and state $f$ to the
higher level state $F$; hence the lower level action $t: i
\rightarrow f$ induces a higher level action $T: I \rightarrow F$. A
\emph{coherent higher level dynamics} $T$ emerges from the lower
level action $t$ if the same higher level action $T$ results for all
lower level states $i$ that correspond to the same higher level
state $I$ \cite{Ell08}, so defining an \emph{equivalence class} of
lower level states that give the same higher level action
\cite{AulEllJae08} (if this is not the case, the lower level
dynamics does not induce a coherent higher level dynamics, as for
example in the case of a chaotic system). Then on coarse graining
(i.e. integrating out fine scale degrees of freedom), the lower
level action results in an emergent higher level dynamics: the
effective theory at the higher level.

\subsubsection{Top-down effects}\label{topdown}

Once higher level structures have emerged, they then exert a
top-down influence on their components (`whole-part constraint') by
constraining the lower level dynamics \cite{Ell11,Ell11a}. In
addition to bottom-up influences, \emph{contextual effects} occur
whereby the upper levels influence what occurs at lower level by
setting the context and boundary conditions for the lower level
actions.\\

This can happen through setting boundary conditions or effective
potentials for the relevant variables (for example creating
electronic band structures that determine how electrons flow in a
solid), or by constraining lower level dynamics through structural
relations (such as the wiring in a computer or synaptic connections
in a brain). This underlies the emergence of effective same level laws of behavior
at higher levels (as in Diagram 1), enabling one to talk of existence of
higher level entities in their own right \cite{Ell08}. It enables true complexity to
emerge through enabling feedback loops between higher and lower levels. \\

The key feature underlying topdown causation is the \emph{multiple
realizability of higher level states} \cite{Ell11}. In a gas, many
lower level molecular states $s_i$ correspond to a specific higher
level state ${\cal S}$ characterized by a temperature $T$, volume
$V$, and pressure $p$. These are the effective macroscopic
variables; one can ordinarily only access the gas by manipulating
higher level variables, hence one cannot determine which specific
lower level state $s_i$ realizes the chosen higher level state
${\cal S}$. It does not matter which specific lower level realizes
the higher level state, what matters is the equivalence class it
belongs to; that is the real causally effective variable
\cite{AulEllJae08}. The number of lower level states that correspond
to a specific higher level state determines the entropy
of that state \cite{Pen11}.\\

It should be emphasized here that these relations can occur between
any two neighboring levels in the hierarchy; there may or may not be
a highest or lowest level.  In \cite{Ell11a} I made the case that
\emph{top-down influences play a key role in the way quantum theory
works}, particularly as regards both decoherence and state
preparation. This paper makes the case that top-down influences are
also key as regards the arrow of time.

\subsubsection{Adaptive Selection}\label{adapt} An important case of top down
causation is adaptive selection \cite{Kau93,Gel94}. Here, selection
takes place from an ensemble of initial states to produce a
restricted set of final states that satisfy some selection
criterion. Random variation influences the outcome by providing a
suite of states from which selection is made in the context of both
the selection criteria and the current environment \cite{Hol92}.
This is the basic process whereby information that is relevant
in a specific context \cite{Roe05} is selected from a jumble
of irrelevant stuff; the rest is discarded. This enables an
apparent local violation of the second law of thermodynamics, as in
the case of Maxwell's Demon (\cite{FeyLeiSan63}:46-5,
\cite{LefRex90}, \cite{AhaRoh05}:4-6; \cite{Car10}:186-189, 196-199)
-- who is indeed an adaptive selection agent, acting against local
entropy growth by selecting high-energy molecules from a stream with
random velocities approaching a trap-door between two compartments
(Figure 1c). The selection criterion is the threshold velocity $v_c$
deciding if a molecule will be admitted into the other partition or
not.
In quantum physics, such a process underlies state vector preparation \cite{Ish95,Ell11a}.\\

-----------------------------------------------------------------------------------\\
\begin{figure}
\begin{center}
 \label{FIGURE 1}
 \includegraphics[width=7.0in]{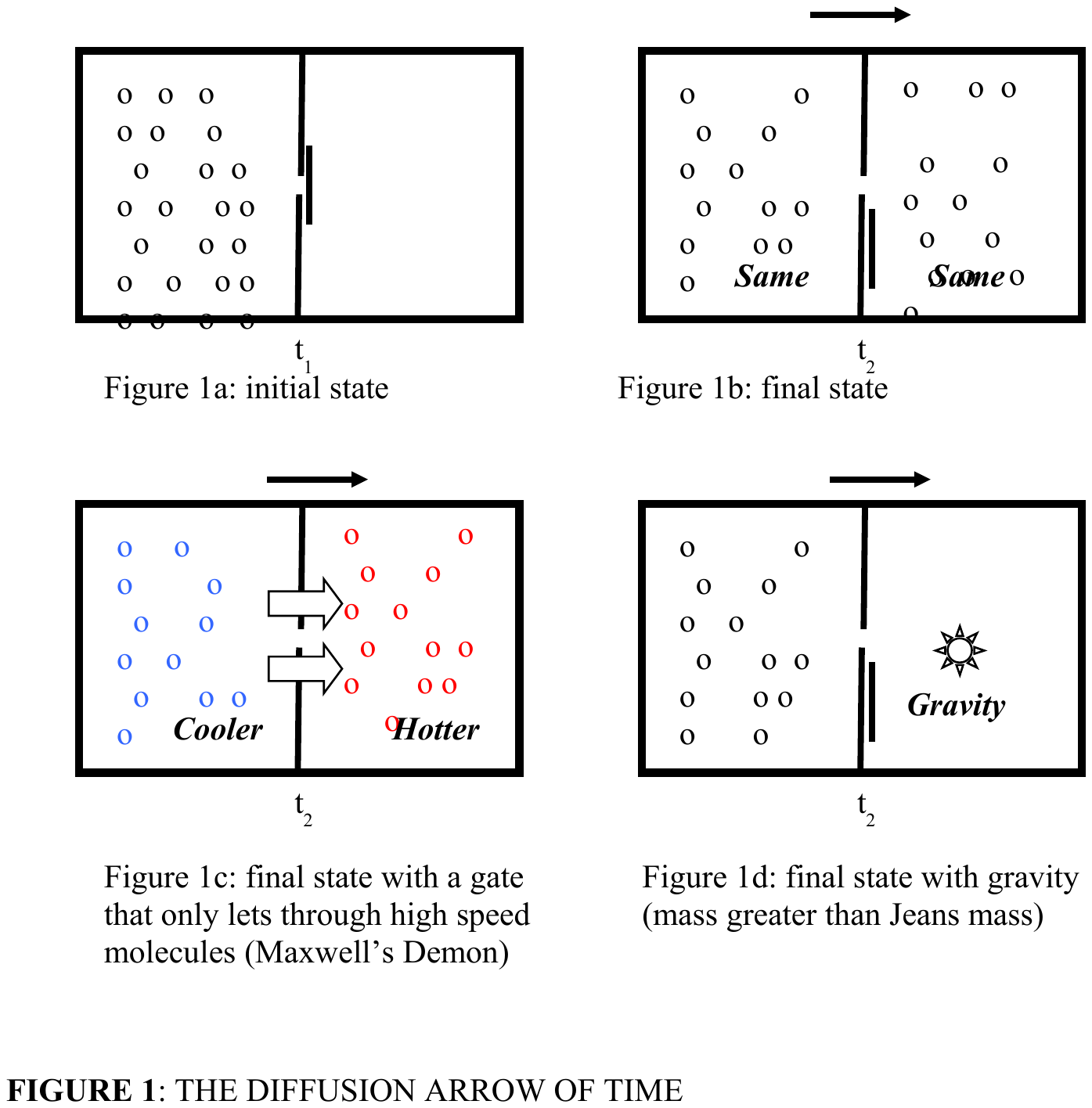}
\end{center}
\end{figure} 

FIGURE 1 HERE\\

Caption: THE DIFFUSION ARROW OF TIME: \\

\texttt{ {\small \textbf{Figure 1a}: Two compartments are separated by an opening,
but it is closed by a slider. Gas is on the left, the right is empty. This is
the prepared starting state: it does not occur naturally.}}

\texttt{ {\small \textbf{Figure 1b}: Opening the aperture results in the gas
spontaneously spreading to the other half, until equilibrium is
reached and entropy is maximized. The arrow of time is evident in
this flow, which is enabled by billions of state vector reduction
events at the quantum level (because the outcome is a well
determined classical state). The transition from the first state to
the second state can be used as an arrow of time detector (given
photographs of state 1 and state 2, you can reliably time order
them). That deduction assumes no human intervention (in fact the
improbable initial state was prepared by humans: another non-unitary
transformation).}}

\texttt{ {\small \textbf{Figure 1c}: If a gate only lets through
high speed molecules, the second law compartment will become hotter
than the first, in apparent contradiction with the second law. This
is an example of creating order by a selection process (slower
molecules are rejected).}}

 \texttt{{\small \textbf{Figure 1d}: If gravity is turned on and the
Jeans mass is attained in the right hand compartment, structure will
spontaneously form. This is presumably in accord with the second law
of thermodynamics, but we have no definition of gravitational
entropy that makes this good. }}\\ 

-----------------------------------------------------------------------------------

\subsection{The central proposal}\label{main1}
Following \cite{Ell11}, a basic standpoint will be adopted (Section
\ref{sec:basis}) and then a central proposal made as to the way the
differen levels relate to each other (Section \ref{prop}).

\subsubsection{A Basic Standpoint}\label{sec:basis}
I adopt the following starting point for what follows:\\

 \textbf{BASIC PREMISE}: \emph{At the classical macroscopic level, Individual Events Happen}. \\

 \noindent Each aspect is important:\\

\textbf{Individual}: Statistics is not enough. An ensemble of events
is made up of individual events. There is no ensemble if individual
events don't separately happen.

\textbf{Events}: Specific things occur. Universal laws describe
multifold possibilities of what might happen, but we experience
specific events in our own particular history.

\textbf{Happen}: They occur in time: they are about to occur, they
occur, then they have occurred. Uncertainty about what might occur
changes to the certainty of what has occurred.\\

\noindent Any theory adopted must recognize that this is the case.
Theories that deal only with statistics of what happens are
incomplete. How this classical behavior emerges from the underlying
quantum theory is of course in dispute; decades of non-realist
interpretations of quantum mechanics claim this is not the way things are  in
the micro realm \cite{Ish95}. However whatever the case is at the
lower levels, in order to be consistent with a huge amount of data,
including our ability to successfully perform experiments, there has
to be some process which leads to emergence of macro realism where this basic premise is true,
whether we understand that process of emergence or not.\\

I will take a particular position on this, associating such
emergence of a classical realm with wave function collapse \cite{Pen89}, but
recognizing there are other proposals that may work. Many of the
considerations that follow will be unchanged whatever that process is.

\subsubsection{Proposal: Nature of physical reality}\label{prop}
The view proposed in \cite{Ell11a} on the nature of physical reality
is as follows.
\begin{enumerate}
  \item \emph{\textbf{Combinatorial structure}: Physical reality is made of linearly behaving components
  combined in non-linear ways}.
  \item \emph{\textbf{Emergence}: Higher level behavior emerges from this lower level structure}.
  \item \emph{\textbf{Contextuality}: The way the lower level elements behaves depends on the
  context in which they are imbedded}.
  \item \emph{\textbf{Quantum Foundations}: Quantum theory is the universal foundation of what happens,
through applying locally to the lower level (very small scale)
entities at all times and places.} \item \noindent \textbf{Quantum
limitations}: \emph{Linearity at higher (larger scale) levels cannot
be assumed, it will be true only if it can be shown to emerge from
the specific combination of lower level elements}.
\end{enumerate}
The same view will be adopted here.

\section{The arrow of time}\label{time2}
A key issue for fundamental physics  is the determination of the
arrow of time.  Section \ref{issue} explains the problem,  and
Section \ref{resolve} considers basic approaches to its resolution:
by coarse graining (Section  \ref{time7_2}), by statistical
fluctuations (Section  \ref{time7_1}), by a foundational quantum
arrow of time (Section  \ref{time7}), and by special initial
conditions (Section  \ref{time7_4}). The latter seems the viable way
to go, and section Section  \ref{time9} develops it in terms of  the
\emph{Past Hypothesis} -- the idea that global conditions determine
the arrow of time by top-down causation. Three possible
Interpretations of this idea are distinguished, and then pursued in
the subsequent sections.

\subsection{The issue}\label{issue}
A crucial aspect of the relation between macro and micro physics is
the origin of the arrow of time (\cite{Edd28}:68-80) : one of the
major puzzles in physics. There is a profound disjunction between
macro and micro physics in this regard.\\

\textbf{At the macroscopic scale}, the Second Law of Thermodynamics
is an unavoidable physical reality \cite{Edd28}: the entropy $S$ of
isolated systems increases with time:
\begin{equation}
dS/dt \geq 0,
\end{equation}
with equality only in equilibrium cases. Irreversibility
relentlessly follows. Examples of irreversibility are
\begin{itemize}
  \item gas in one half of a container spreading to fill the whole
  when a partition separating it from the other half is removed (Figures 1(a), 1(b)),
  \item a glass falling off a table and smashing to pieces \cite{Pen89,Pen11},
\item water flows downhill,
  \item a block sliding on a plane and coming to a stop owing to
friction,
\item a stone tossed into a lake and sending out waves
along the surface of the water,
 \item a radio signal or sound wave is received after it was sent,
\item a footprint left in the sea sand after you have walked past,
\item the moving finger writes and moves on, leaving its trace behind (Omar Khayam),
\item the progress of life from birth to death: the seven ages of mankind (Shakespeare).
\item the evolution of life on earth: once there was no life, now
there is;
\item the progression of structure growth in the universe: once
there was no structure; now there is.
\end{itemize}
Thus the arrow of time and irreversibility of physical effect occurs
on all scales of the hierarchy, except perhaps the quantum scale;
but it occurs there too if one accepts the reality of wave function
collapse (\ref{trans}).\\

 This irreversibility relates to
\textbf{loss of useable energy} as the passage of time occurs, and
associated \textbf{increase of disorder} (\cite{Car10}:143-171). It
is a core feature of thermodynamics \cite{Fuc96} and physical
chemistry \cite{Atk94}, and hence plays a crucial role in biology
(\cite{CamRee05}:143-144), energy flows in ecosystems
\cite{Sch05,Wik11},
and energy needs of an industrial economy \cite{Geo71}. \\

\textbf{At a microscopic level}, with one caveat I attend to
shortly, the basic interaction equations for the four fundamental
forces are time symmetric, and so coarse graining them should lead
to time symmetric macroscopic laws. The unitary evolution described
by the quantum evolution equations also does not determine a
direction of time, because the underlying unitary theory treats the
future and past directions of time as equal. Specifically: in
equation (\ref{U2}), one can make the swap $t_1 \leftrightarrow t_2$
and get an identical solution to (\ref{U2}), but with the opposite
arrow of time (let $t \mapsto \tilde{t}: = -t$ and the solution will be
identical to (\ref{U2})). The same is famously true of Feynman diagrams \cite{Fey48}.

\subsubsection{The micro-macro relation}
Macro effective laws are
often determined by coarse-graining micro laws (Section \ref{emergeh}).
\emph{The macro laws that emerge by coarse graining should have the
same time symmetry as the micro laws} (simply reverse the arrow of
time in the coarse graining process in Diagram 1). This is true even
when we deduce higher level equations for irreversible statistical
behavior: there will be an equally good solution with the opposite
direction of time. Hence there is apparently a fundamental
contradiction:
\begin{quote}
\emph{The macro behavior displays a time asymmetry that is not
apparent in the fundamental equations out of which they emerge.}
\end{quote}

\noindent Thus for each solution of the equations of Newtonian
dynamics, of Newtonian gravity, of electromagnetic theory, of
special relativity there is a time reversed solution of the
equations where everything happens in the opposite sense of time. In
the case of the glass falling off a table and smashing to pieces,
there is a time reversed solution where the pieces of the glass
assemble themselves into a whole glass and ascend back onto the
table \cite{Pen89}. In the case of the water waves, there is a time
reversed solution where spherical incoming waves converge on a point
and pull the stone back
up out of the water \cite{Zeh07}. But we never see this happen in practice.\\

\textbf{The basic problem}: \emph{\textbf{How does the macro theory
determine which is the future as opposed to the past, when this time
asymmetry is not apparent in the underlying unitary theory}}?
\cite{EllSci72,Dav74,Zeh07,HalPerZur96,Cal11}. \\

The caveat mentioned above is that there is a very weak time
asymmetry of weak interactions. However this seems too ineffective
to be the origin of the time asymmetry we see at macroscales: the
weak interaction does not have enough purchase on the rest of
physics (indeed the time asymmetry is very difficult to detect).

\subsection{Possible resolutions}\label{resolve}
This Section considers in turn resolution by coarse graining,   by
statistical fluctuations, by a foundational quantum arrow of time,
and by special initial conditions. The first three are bottom-up
approaches that all cannot succeed because of Loschmidt's paradox
(Section \ref{time7_2}). Hence a top down approach - special
cosmological initial conditions -- has to be the way to go (Section
\ref{time7_4}).

\subsubsection{Bottom-up resolution by coarse graining }\label{time7_2}
Now an initial reaction is that coarse graining from micro to macro
scales results in an arrow of time, as shown beautifully by
Boltzmann's H-Theorem (\cite{Zeh07}:43-48), resulting from the fact
that random motions in phase space takes one from less probable to
more probable regions of phase space (\cite{Pen04}:686-696;
\cite{GemMicMah04}:43-47; \cite{Car10}:172-174) \cite{Pen11}:9-56).
Hence one can show that entropy increases to the future; the second
law of thermodynamics at the macro level emerges from the coarse
grained underlying micro theory. The quantum theory version of this
result is the statement that the density matrix open system evolves
in a time asymmetric manner, leading to an increase in entropy
(\cite{BrePet06}:123-125).\\

But this apparent appearance of an arrow of time from the underlying
theory is an illusion, as the underlying theory is time symmetric,
so there is no way an arrow of time can emerge by any local coarse
graining procedure. Indeed the derivation of the increase of entropy
in Boltzmann's H-Theorem applies equally to both directions of time
(swap $t \rightarrow -t$, the same derivation still holds). The same applies
to any derivation from quantum field theory, for example that given by Weinberg \cite{Wei95}.\\

This is Loschmidt's paradox (\cite{Pen89}: Fig 7.6;
\cite{Pen04}:696-699; \cite{Pen11}):
\begin{quote}
\emph{The H-theorem predicts entropy will increase to both the
future and the past}.
\end{quote}
(Figure 2). The same will apply to the quantum theory derivation of
an increase of entropy through evolution of the density matrix
(\cite{BrePet06}:123-125, \cite{GemMicMah04}:38-42, 53-58): it
cannot resolve where the arrow of time comes from, or indeed why it
is the same everywhere.\\

-----------------------------------------------------------------------------------\\
\begin{figure}
\begin{center}
\label{FIGURE 2}
 \includegraphics[width=7.0in]{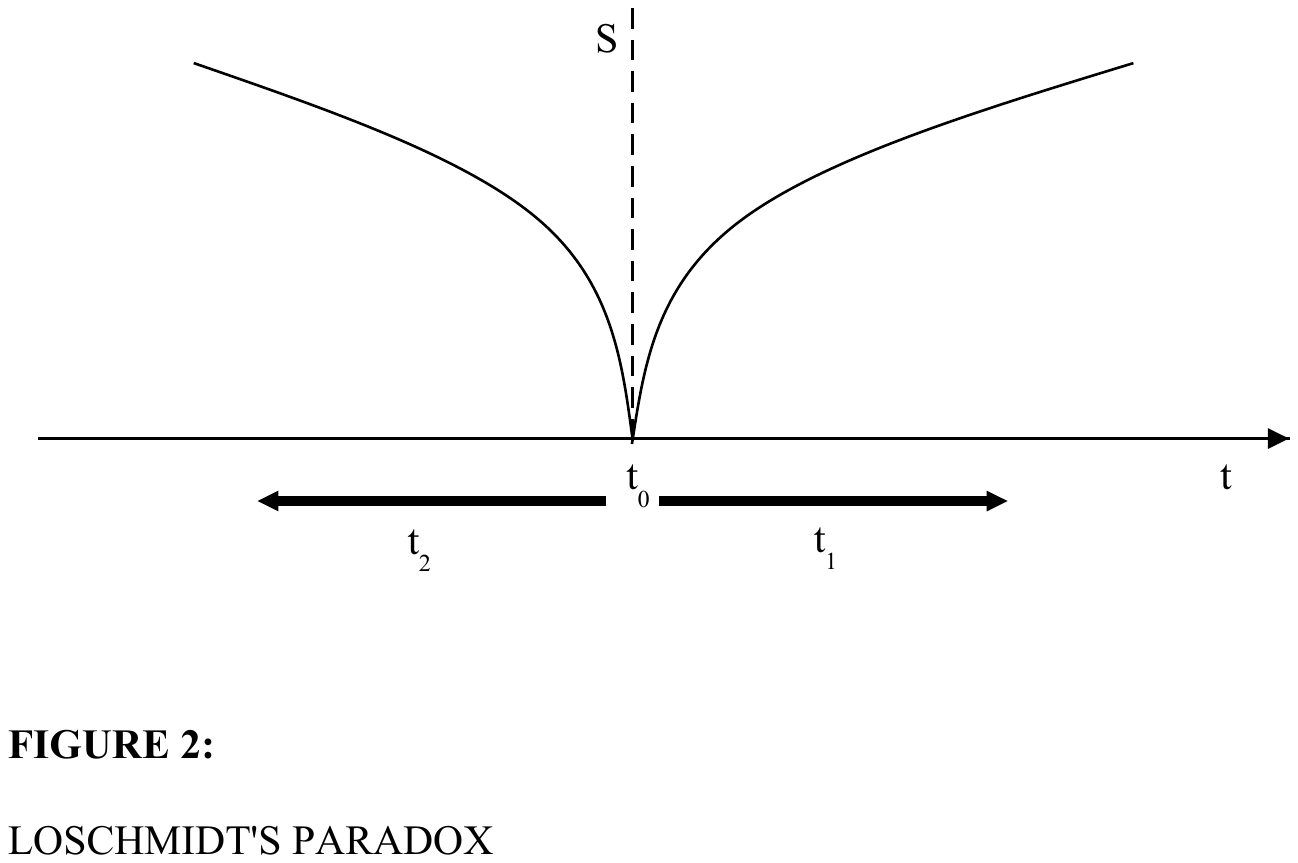}
\end{center}
\end{figure}

FIGURE 2 HERE\\

Caption:  LOSCHMIDT'S PARADOX: \\

\texttt{{\small Set initial conditions for a system time $t_0$. Let time
physics evolve in the future direction of the time coordinate $t$, giving time $t_1$. Entropy $S(t)$
increases (Boltzmann's H-Theorem), so $S(t_1) \geq S(t_0)$. But this does not account for the fact that the
initial direction of time $t$ was chosen arbitrarily. The dynamics is time symmetric; it could equally set off in
the opposite direction, giving development to time $t_2$. Boltzmann's H-Theorem applies equally in this direction
(set $t \mapsto -t$, the proof is unchanged), implying $S(t_2) \geq S(t_0)$. This is
Loschmidt's paradox: the H-theorem works in both direction of time \cite{Pen89}.}}\\

-----------------------------------------------------------------------------------

\subsubsection{The locality issue}\label{sec:cg}
The latter is a key question for any local proposal for determining
the arrow of time:
\begin{quote}
\textbf{The arrow of time locality issue}:\emph{ If there is a
purely local process for determining the arrow
  of time, why does it give the same result everywhere}?
\end{quote}
Local determination has to arbitrarily choose one of the two
directions of time as the positive direction indicating the future;
but as this decision is made locally, there is no reason whatever why
it should be consistent globally. if it emerges locally, opposite arrows may be expected to
occur in different places.\\

We are unaware of any contradictions as regards the direction of the
arrow of time, either locally (time does not run backwards anywhere
on Earth) or astronomically (irreversible process in distant
galaxies seem to run in the same direction of time as here
\cite{Ree95}). Some coordinating mechanism is called for to ensure the arrow of time points in the same
direction everywhere in our past.

\subsubsection{Bottom-up resolution by statistical fluctuations}\label{time7_1}
It is often claimed that if one has an equilibrium state that lasts
an infinite time, fluctuations round equilibrium can lead to any
state whatever popping out of the vacuum just as a statistical
fluctuation, with associated emergence of a local arrow of time. This leads to
Poincare's \emph{Eternal return} (any state whatever that has
occurred will eventually recur) and the \emph{Boltzmann Brain}
scenario: you can explain the existence of Boltzmann's brain not as
a result of evolution but just as an eventual inevitable result of
statistical fluctuations if an infinite amount of time is available
(\cite{Car10}:201-227). \\

This is problematic in the real universe, as we have not had 
statistical equilibrium anywhere except for very short timescales in
very local contexts since the end of inflation. And it would also run into the locality issue
(Section \ref{sec:cg}): such fluctuations would be local in space,
and different arrows of time could emerge in different places. The
context for relevance of these arguments has not occurred in the
real universe: the argument does not take realistic context into
account (\cite{FeyLeiSan63}:46;\cite{Zeh07}:42), except possibly in
the far future of the universe if it expands forever due to a
cosmological constant (\cite{Car10}:313-314). One cannot explain the
arrow of time we experience at the present moment as being a consequence of statistical fluctuations.


\subsubsection{Bottom up resolution by a foundational quantum arrow of time}\label{time7}
 Could a resolution come from the local time asymmetry in state vector projection
 (\ref{trans})? After all if quantum physics underlies all the rest, perhaps
 the time asymmetry involved in
(\ref{trans}) could be the source for the rest, based in the way a
local emergence of classicality works through collapse of the wave
function (\cite{Pen04}:527-530; \cite{AhaRoh05} 136-147) and
associated increase of entropy \cite{Zeh90}. But then where does
that quantum level time asymmetry come from? That depends on the
resolution of the unresolved issue of state vector reduction. I will
make the case that the nature of this process is largely determined by local
top-down effects (Section \ref{topdown}) due to the specific nature
of local physical structures \cite{Ell11a}. In that case, this
asymmetry may be determined locally by top-down causation, rather
then being the source of the asymmetry. This conclusion is
reinforced by the locality issue (Section \ref{sec:cg}): in order to
assign an arrow of time everywhere in a consistent way, it has to be
determined contextually through some coordinating
mechanism ensuring it is the same everywhere in a connected spacetime domain.\\

The likely solution is that resolution is by a  top-down effect, the
local time asymmetry of state vector reduction being based in a time
asymmetry in the local detection environment, in turn founded in
conditions in the universe as a whole: the local environment too has
to know which time direction to choose as the future, else the set
of local environments too will fall foul of the locality issue. The
issue arises for each level in the hierarchy of complexity (Table
1): if it is not determined from below, it must be determined from
above. Considering higher and higher levels, the answer must lie at
the top.

\subsubsection{Resolution by large scale initial conditions}\label{time7_4}
The implication must be that the arrow of time results from global
environmental conditions, as it can't reliably emerge in a
consistent way from local physics that does not care about the
direction of time. Feynman stated in his lectures,

\begin{quote}
``\emph{So far as we know all the fundamental laws of physics, like
Newton's equations, are reversible. Then where does irreversibility
come form? It comes from going from order to disorder, but we do not
understand this till we know the origin of the order... for some
reason the universe at one time had a very low entropy for its
energy content, and since then the entropy has increased. So that is
the way towards the future. That is the origin of all
irreversibility, that is what makes the process of growth and decay,
that makes us remember the past and not the future}...''
(\cite{FeyLeiSan63}:46-8)
\end{quote}
This fits into the fundamental nature of causality in the following
way: a key feature of causality as determined by physical equations
is that (\cite{Har03};\cite{Zeh07}:1-3) the outcome depends both
on the equations plus the initial and/or final conditions. Hence
\begin{quote}
\emph{\textbf{Broken symmetry}: The solution of a set of equations
will usually not exhibit the symmetry of the underlying theory} \cite{And72}. \emph{If
there is no time asymmetry in the equations, it must lie in the
initial and/or final conditions}.
\end{quote}
Note that precisely because we are dealing with time asymmetry, we
cannot assume it is the initial conditions alone; in principle we may need to compare them to final conditions (\cite{WheFey45};
\cite{Pen89}). However I will make the case below that we need to be
concerned only with initial conditions, because the relevant context is that of an Evolving Block Universe (Section \ref{EBU}).

\subsection{Top-down determination: The Past Hypothesis}\label{time9}

 Thus the only viable option seems to be the \emph{Past Hypothesis} (\cite{Alb00};
\cite{Car10}:176):
\begin{quote}
 \emph{\textbf{The direction of time
must be derived by a top-down process from cosmological to local
scales}}.
\end{quote}
It is strongly supported by the fact that the entropy of universe
could have been much larger than it was (\cite{Pen89};
\cite{Car10}:345-346) because black holes could have had much more
entropy (\cite{Car10}:299-302; \cite{Pen04}:728-731)). It started
off in a very special state, characterized by the \emph{Weyl Curvature
Hypothesis} (the universe is asymptotically conformally flat at the big bang
(\cite{Pen04}:765-769)), which was required in
order that inflation could start \cite{Pen89a}.\\

\noindent To investigate this further, I distinguish the following
three possible aspects:
\begin{itemize}
  \item \textbf{AT1: Global time asymmetry}:  \emph{a difference in conditions at the start and end of the universe}
(\cite{WheFey45}, \cite{FeyLeiSan64}:28-6, \cite{EllSci72});
  \item \textbf{AT2: Global past condition}: \emph{Special conditions at the start, on cosmological scale:
  the expanding universe started in a special low entropy condition},\footnote{Arrow of time
arguments are notoriously tricky. If you consider the arrow going
the other way, then this state would be the end, not the start.}
\emph{which thus made it possible for it to evolve towards higher entropy
states} (\cite{Pen04}:702-707, \cite{Car10}, \cite{Pen11}:57-136)) \emph{and solves Loschmidt's paradox because the global past condition cascaded down to give a sequence
of local past conditions.}
  \item \textbf{AT3: An initial master arrow of time}: \emph{the other arrows derive from the global master arrow of
  time resulting from the universe's early expansion from an initial singularity in an Evolving Block Universe} \cite{Ell13}.
  \emph{The arrow of time at the start is the time direction pointing away from the initial singularity towards the growing
  boundary of spacetime; this then remains the direction of time at all later times.}
\end{itemize}

\noindent Such cosmological asymmetries provide a possible source
determining why the local arrow of time is the way it is, by
top-down causation from the global to the local direction of time
(the latter will therefore be the same everywhere, avoiding the
arrow of time locality problem). \\

How are these different reasons related to each other? It might seem
that \textbf{AT3} might be reducible to \textbf{AT2}; but this is not the case. \textbf{AT3} is
defined in the context of an evolving block universe \cite{Ell13},
where the flow of time would be determined as the time direction
leading away from the in initial singularity, pointing from the
start to the growing edge of spacetime, and so providing the `master
arrow',which would cascade down to the other arrows as discussed
below. If the start of the universe occurred in a very inhomogeneous
way, \textbf{AT2} would not be satisfied but \textbf{AT3} would still set the direction of time. If the singularity was
inhomogeneous enough, the Second Law would not hold (entropy would
not increase in the future direction of time).\\

I will make the case that \textbf{AT1} is not the way to go, because
the proper context to view the situation is an evolving block
universe \cite{Ell13} (see Section \ref{sec:time1}), which rules
this proposal out. Rather the direction of the flow of time is due
\textbf{AT3}, which provides the master direction of time. Then
\textbf{AT2} is required in order that entropy increase as time
flows. As Loschmidt's paradox makes clear, entropy can in principle
increase in either direction of time; the master arrow \textbf{AT3}
makes the choice as to which direction is the future, while
\textbf{AT2} makes sure entropy increase follows that arrow. The
other arrows of time \cite{EllSci72,Dav74,Zeh07,HalPerZur96}
(thermodynamic, electrodynamic, gravitational, quantum, and
biological) all then follow.\\

In order to make this precise, I will distinguish between the
\emph{direction of time} and the \emph{arrow of time} as follows:
\begin{quote}
\textbf{The direction of time} is the cosmologically determined
direction in which time flows globally. It represents the way
spacetime is continuously increasing as an Evolving Block Universe.
\end{quote}
By contrast,
\begin{quote}
\textbf{The arrow of time} is the locally determined direction in
which time flows at any time in the evolution of the universe. It
represents the way physics and biology manifest the flow of time
locally.
\end{quote}

-----------------------------------------------------------------------------------\\
\begin{figure}
\begin{center}
\label{FIGURE 6}
 \includegraphics[width=7.0in]{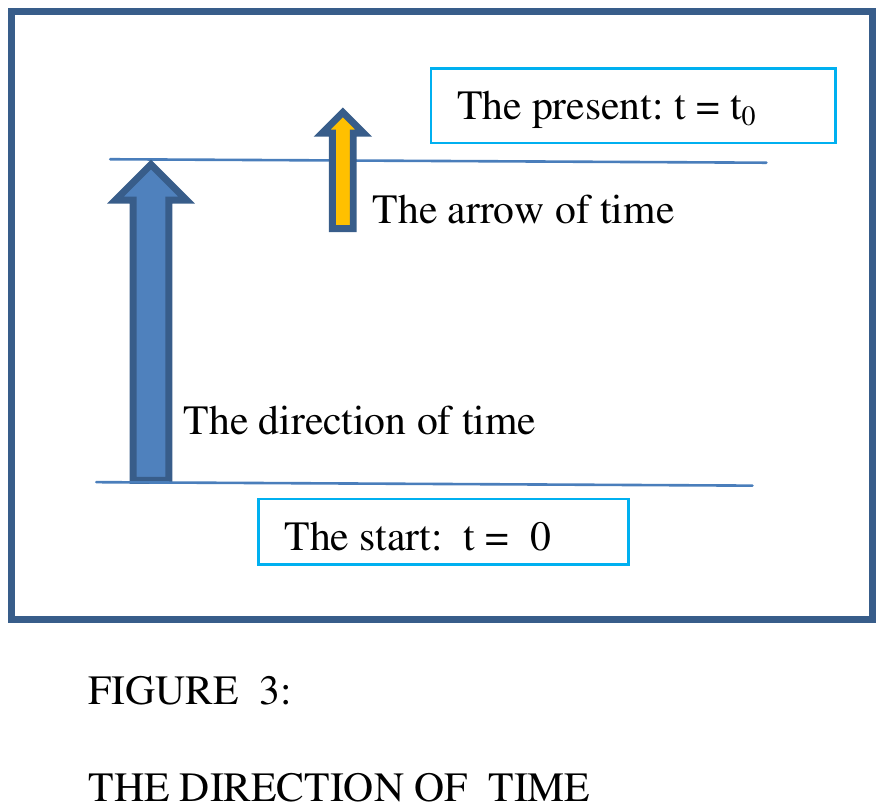}
\end{center}
\end{figure}

FIGURE 3 HERE\\

Caption:  THE DIRECTION OF TIME: \\

\texttt{{\small The direction of time is the cosmological direction of time from the start of the universe to the present. It corresponds to the direction on which the Evolving Block universe is growing. The arrow of time is the local direction of time affecting local processes, so it is the direction in which in which entropy is increasing.  The arrow of time at each time devolves from the global direction of time.}}\\

-----------------------------------------------------------------------------------\\

The proposal will be that,
\begin{itemize}
  \item the flow of time, and hence the direction of time, is determined by
the cosmological master arrow of time \textbf{AT3};
  \item this then determines the arrow of time for local physical processes by a top-down
cascade in the hierarchy of physical structure, based on special
cosmological initial conditions \textbf{AT2};
  \item this in turn determines the arrow of time for complex systems and life by a
bottom-up cascade in emergent systems.
\end{itemize}

\section{The start and continuation of time}\label{time3} To look
at this, we must have a reasonable model of cosmology from some
starting time through structure formation up to the present day
\cite{Har00,Sil01,Dod03,EllMaaMac11}. Section \ref{cosm time} shows
how cosmic time is set up, Section \ref{epochs} discusses the main
relevant cosmic epochs since the start of inflation, and Section
\ref{time11} the speculative pre-inflation possibilities.

\subsection{Cosmic time}\label{cosm time} The background model
used in cosmological studies is the
Friedmann-Lema\^{\i}tre-Robertson-Walker (FLRW) spacetime, given
in comoving coordinates by
\begin{equation}\label{FLRW}
ds^2 = - dt^2 + a^2(t) d\sigma^2
\end{equation}
where $d\sigma^2$ is a 3-space of constant curvature and $a(t)$ the
scale factor \cite{HawEll73,Ell06a,EllMaaMac11}. Perturbations
around that model characterize how structure formation took place
\cite{Dod03,EllMaaMac11}.\\

 The dynamics of the FLRW model is governed by three
 interrelated equations. The
 \textit{energy-density conservation equation} determines the time
 evolution of the density $\rho(t)$:
\begin{equation}
\dot{\rho}+(\rho +p/c^{2})3\frac{\dot{a}}{a}=0\,. \label{cons}
\end{equation}
and so determines the evolution of the pressure $p(t)$
 through a suitable equation of state $p = p(\rho)$.
 Second, the scale factor $a(t)$ obeys the \emph{Raychaudhuri
equation}
\begin{equation}
3\frac{\ddot{a}}{a}=-\frac{1}{2}\kappa (\rho +3p/c^{2})+\Lambda ,
\label{Ray}
\end{equation}
where $\kappa $ is the gravitational constant and $\Lambda $ the
cosmological constant. Third, the first integral of equations
(\ref{cons}, \ref{Ray}) when $\dot{a}\neq 0$ is the \emph{Friedmann
equation}
\begin{equation}
\frac{\dot{a}^{2}}{a^{2}}=\frac{\kappa \rho}{3}+\frac{\Lambda
}{3}-\frac{k}{ a^{2}}.  \label{Fried}
\end{equation}
where $k$ is an integration constant related to the spatial
curvature. Thus the cosmic time is the time parameter $t$ that
enters into these equations determining the scale factor evolution.
It is determined by being the time parameter naturally appearing in
the 1+3 covariant formulation of the Einstein Field equations in the
cosmological context \cite{EllMaaMac11}, which reduce to equations
(\ref{cons})-(\ref{Fried}) when specialized to a FLRW geometry. But
the equations (\ref{cons})-(\ref{Fried}) are invariant
under $t \rightarrow \tilde{t} := -t$: if $a(t)$ is a solution, so is if $a(\tilde{t})$.\\

Classically, the universe began at a spacetime singularity
\cite{HawEll73}, conventionally set to be $t=0$. Cosmic time starts
at the creation of universe: time came into being, it did not exist
before (insofar as that makes sense). At any small time $t =
\varepsilon >0$, the arrow of time is defined to point from the
singularity at $t=0$ to the present time $t = \varepsilon$ (the
future boundary of the evolving spacetime), because that is the
direction of time in which spacetime is increasing in the evolving
block universe \cite{Ell13}. Therefore
\begin{quote}
\textbf{The Direction of Time}: \emph{if the proper time $\tau_P$
along a fundamental world line from the initial singularity to the
event $P$ is greater than the proper time $\tau_Q$ along that world
line from the initial singularity to the event $Q$, then the
direction of time is from $Q$ to $P$.}
\end{quote}
 After it has come into being there is no way it can
reverse, because once spacetime has come into existence, it can't
disappear. Once the flow of time is established it just keeps
rolling along, determining what happens according to
(\ref{cons})-(\ref{Fried}) unless we reach a spacetime singularity
in the future, when it comes to an end.
\begin{quote}
\emph{\textbf{The cosmological direction of time \textbf{AT3} is set by the start
of the universe.} There is no mechanism that can stop or reverse the
cosmological flow of time, set by the start of the universe. It sets
the direction of flow of the time parameter $t$ in the metric
(\ref{FLRW}); time starts at $t=0$ and then increases monotonically
along fundamental world lines, being the parameter $t$ occurring in
the solution $\{a(t), \rho(t), p(t)\}$ to equations
}(\ref{cons})-(\ref{Fried}).\footnote{If we were perverse we could
use the reverse time label where time proceeded from $t = 0$ to
values $t<0$; but this is just a coordinate convention with no
effect on the physics. It is psychologically sensible to assign the
sign so that $t$ proceeds to positive values. This sets the
convention for the concepts `later' and `earlier' in the usual way.}
\end{quote}
This is the master arrow of time \textbf{AT3}. The gravitational
equations (\ref{cons})-(\ref{Fried}) are time symmetric (because the
Einstein equations are time symmetric), but the actual universe had
a start. This broke the time symmetry and set the master arrow of
time: the universe is expanding, not contracting, because it started off from a zero volume state. It had nowhere to grow but larger.\\

How this evolution actually occurs is determined by the changing
equation of state of the universe at different epochs (next
section). When we take account of quantum gravity, this picture is
altered, and various options arise (\cite{Ell04}:16-18); these are
discussed below in Section \ref{time11}. However the universe will
still set a unique monotonically increasing time for all epochs
after the end of the quantum gravity epoch.

\subsection{The cosmic epochs}\label{epochs} The basic dynamics
of cosmology to the present time can be regarded as having five
phases (\cite{Dod03}:1-20),\footnote{See \cite{Ell06a}: Sections
2.1-2.2, 2.6-2.8 for a conveniently accessible short description.}
summarized in Figure 3:
\begin{itemize}
  \item \textbf{Epoch 0: Pre-Inflationary era.} Any quantum
  gravity era that might precede inflation. The dynamics at this
  time is hypothetical: we don't know what happened then (there may or may not have
  been an actual physical start to the universe).
  \item \textbf{Epoch 1: Inflationary era.} A very brief period of exponential expansion,
  ending at reheating and conversion of the inflaton field to
  radiation, marking the start of the Hot Big Bang era. Inflation is the
  time when quantum perturbations arose that provided the seeds for
  structure formation at the end of the Hot Big Bang era.
  \item \textbf{Epoch 2: Hot Bang era.} An epoch of radiation and matter in
  quasi-equilibrium, up to the time of decoupling of matter and
  radiation at the \emph{Last Scattering Surface}
  (`LSS'). This epoch includes baryosynthesis, nucleosynthesis,
  and the transition from a radiation dominated to matter dominated
  expansion. The universe was opaque up to the end of this era.
  \item \textbf{Epoch 3: Structure formation era.} The epoch from the LSS to
  the present day. The universe became transparent, matter and
  radiation decoupled leading to the universe being permeated by
  Cosmic Black Body Radiation (CBR), and structure formation commenced, leading to
  the existence of large scale structures, galaxies, stars, and
  planets. At a late time (close to the present day) dark energy
  started to dominate the dynamics, leading to a speed-up of the
  expansion of the universe.
  \item \textbf{Epoch 4: The future.} The epoch from the present day on, either
  an unending expansion (the most likely option), or recollapse to a future singularity; which is the case depends on
  parameters and physics that is not well known.
\end{itemize}

-----------------------------------------------------------------------------------\\

\begin{figure}
\begin{center}
\label{FIGURE 3}
 \includegraphics[width=6.0in]{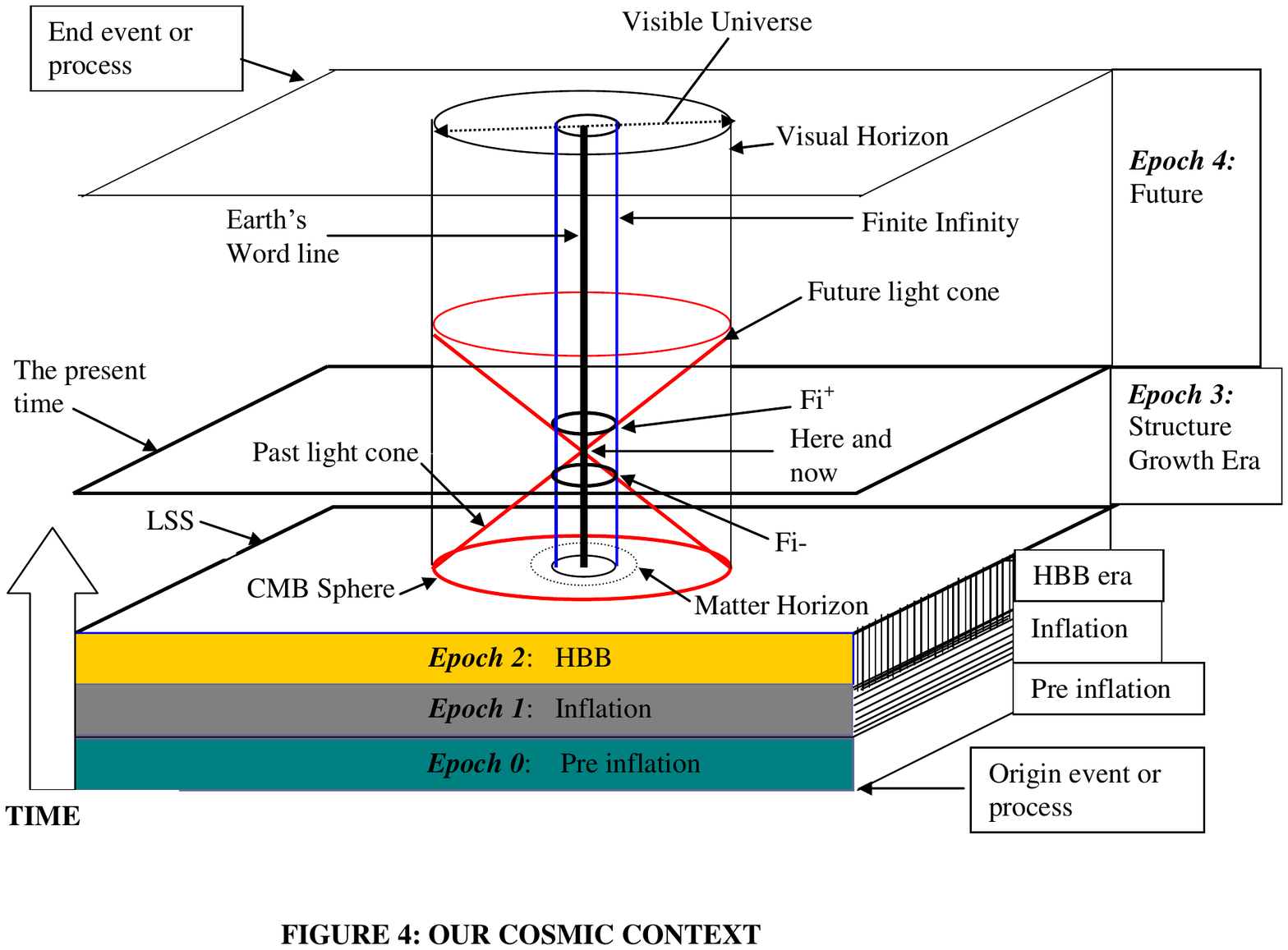}
\end{center}
\end{figure}

FIGURE 4 HERE\\

Caption: OUR COSMIC CONTEXT\\

\texttt{{\small A conformal diagram of the cosmic context for
local existence. Time runs vertically, 2 space dimensions
horizontally (one space dimension is hidden). Light rays travel at
$45^o$ to the vertical. The start of the universe is indicated at
the bottom (this might possibly represent a start a finite time ago,
or at minus infinity; conformal diagrams do not represent distance
or proper time accurately). This is followed by a pre-inflation
quantum gravity era, an inflationary era, and a Hot Big Bang (HBB)
era, which ends at the surface of last scattering (LSS). The LSS
marks the start of structure formation, which extends from the LSS
to the present time. Structure formation may continue for some time
to the future, but will eventually come to an end. The far future
boundary of the universe may lie a finite time to the future, but
more probably is an infinite proper time to the future.}}

\texttt{{\small The Earth's world line is the vertical line at the center, with the present time
``here and now'' marked. Our past light cone extends down to the
LSS, which it intersects in a 2-sphere; this is set of events from
which the 2.7K cosmic microwave background (CMB) originated. We
cannot see to earlier times because the universe was opaque in the
HBB era; hence the matter world lines through this 2-sphere form our
visual horizon (the surface in spacetime separating matter we can
have seen from that which we cannot detect by any electromagnetic
radiation). For all practical purposes, ``infinity'' for local
physics is a sphere of radius 1 light year around  the Earth. This
is our ``Finite Infinity'', its world tube surrounding our
world-line in spacetime. Our future light cone intersects it in the
2-sphere $Fi^+$ (our outgoing radiation sphere), and our past light
cone intersects it in the 2-sphere $Fi^-$ (our incoming radiation
sphere); this is our effective sky -- every star and galaxy we see
is an image on this sphere.}
}
\texttt{{\small On the Block Universe view of spacetime,
everything here from the start to the finish exists as a single
spacetime block, where all times are equal so the present has no
meaning. On the Emerging Block Universe (EBU) view, the present is
the special time where, at this instant, the uncertain future is
changing to the determined past. Hence on this view, the nature of
existence is different in the past (below the surface labeled ``The
present time'') and in the future (above that surface). The former
exists (as it has been determined), whereas the latter is presently
only potential (so does not yet exist). As time progress, the
present time moves up our world line, so the past region of
spacetime is continually getting bigger: spacetime is growing. In
the far future, when everything has happened, the present will
coincide with the future boundary and the EBU will have evolved into
a Final Block
Universe. } 
}\\

-----------------------------------------------------------------------------------\\

The dynamical behavior is different in each epoch.  To a good
approximation we can represent the inflationary era, radiation
dominated era, and matter dominated era as
\begin{eqnarray}
\textbf{Inflation:\,}\,\,& p = \rho c^2 &\Rightarrow \,\,a(t) = \,\,
a_0 \,e^{H (t- t_0)} \label{frw1}\\
\textbf{Radiation dominated:\,}\,\, &p = \frac{1}{3}\rho c^2 &
\Rightarrow \,\, a(t) = \,\, a_1 \,(t-t_1)^{1/2}\label{frw2}\\
\textbf{Matter dominated:\,}\,\, &p = 0\,\, & \Rightarrow \,\, a(t)
= \,\,a_2\, (t-t_2)^{2/3}\label{frw3}
\end{eqnarray}
Dark energy started to be significant late in Epoch 3, altering
(\ref{frw3}) a bit at late times, but it only alters the big picture
significantly in the future era (Epoch 4), where it suggests there
will be no final singularity: equation (\ref{frw1}) will hold again and expansion will last for ever
(but the physics is uncertain: other options are possible).


\subsection{The speculative pre-inflationary era}\label{time11}
What previous mechanism could lead to a universe satisfying the
past condition? Why is the universe an almost-FLRW universe
when it emerges from the inflationary era? What set the conditions before inflation such that inflation could start?\\

We are in a domain of untestable speculation here, and
the possibilities we can imagine certainly won't encompass all that
might have been the case. Still it's fun to speculate. The problem
is to find a way that generates a smooth start to the universe,
given that inflation can't do so for generic initial conditions \cite{Pen89,Pen89a}.
Amongst the possibilities are,
\begin{itemize}
  \item \textbf{The smooth universe machine} Whatever creates the
  universe makes a nice smooth universe: it works in a uniform way
  at all emergent spacetime events. This was the assumption everyone
  made from the 1930s until the 1970s \cite{Har00}.
  \item\textbf{The multiple universe machine} Whatever creates the
  universe keeps doing it, with variation. This includes the chaotic
  inflation scenario. Both arrows of time can occur in different
  domains (\cite{Car10}:359-364, 371-372;\cite{CarTam10}).
  \item\textbf{The bounce machine} The present epoch of the universe
  resulted from some kind of bounce or rebirth from a previous era
  that sets up the special state needed (\cite{Car10}:349-353; \cite{Pen11}).
  \item \textbf{The emergent universe machine} Whatever creates the
  universe makes an emergent universe that spends a long time in an almost
  static state, with compact spatial sections \cite{EllMaa04,EllMurTsa04,Muletal05}.
  The previous state
  does not matter as there is plenty of time for matter to
  come into equilibrium: causal effects can travel round the
  universe thousands of times.
  \item \textbf{The quantum gravity machine} Whatever creates the
  universe does so in a quantum gravity era where causality has not
  yet emerged, for example there is a spacetime foam,
  so causal restrictions don't yet exist. Horizons emerge later. Another option is the
  causal set approach to quantum gravity where spacetime emerges from a discrete basic structure \cite{Hen06}.
  This is in accord with the view put here, as it allows a growing universe like the Evolving Block Universe.
\end{itemize}
Whatever happened in this era is the ultimate source of the arrow of time, but the physics is
completely uncertain, as is the overall context. I will here assume that whatever was needed happened, and
led to a suitable start of inflation.

\section{The descent of time: contextual effects}\label{time4}
The basic idea in this section is that the expansion of universe determines the arrow of time at in the natural sciences
hierarchy so as to concur with the direction of time set by the expansion of the universe \textbf{AT3}. The arrow of time ripples down from higher to lower levels in the physical hierarchy, as a consequence of the special global initial
conditions \textbf{AT2}; this process sets up similar speciality conditions at smaller scales in the hierarchy.\\

Crucially, at the beginning of the HBB era the universe was
expanding and cooling, not vice versa. This sets the arrow of time
for local physics, by passing a variable (the global temperature
$T(t)$ of matter and radiation), determined by the global expansion
history (Section \ref{epochs}) to local physical systems. $T(t)$ is
decreasing as $t$ increases; this is what determines the local arrow
of time at each lower level. It gets communicated to all the lower
level physics because they are systems imbedded in a heat bath with
decreasing temperature. Initially, it is a heat bath created by
being immersed in the expanding cosmic fluid; later after
decoupling, it is a radiative heat bath. Additionally matter is
moving apart and thinning out: so density decreases as time
progresses, with comoving world lines moving
ever further apart, constraining causal processes.\\

This occurs in the inflationary epoch (Section \ref{epoch1}), the
hot big bang era (Section \ref{epoch2}), and the astronomical epoch
(Section \ref{epoch3}). This determines the local thermodynamic
arrow of time (Section \ref{epoch4}) and the radiative arrow of time
(Section \ref{radiate}). To examine the latter clearly, it is
useful to introduce the idea of a finite infinity for isolated
systems (Section \ref{isolated}). Finally the arrow of time cascades
down from local systems to microsystems (Section \ref{epoch_micro}).

\subsection{Epoch 1: Inflation}\label{epoch1} This exponentially
expanding era (\ref{frw1}) is driven by a scalar field called the
inflaton (\cite{Sil01}:115-123). It has three effects:
\begin{itemize}
  \item The exponential expansion causes initial inhomogeneities and curvature to decay
  away (\cite{Dod03}:144-155).
  \item Quantum fluctuations generate tensor perturbations that
  result in gravitational waves (\cite{Dod03}:155-162).
  \item Quantum fluctuations generate scalar perturbations that
  result in density inhomogeneities that later on are the seeds
  for large scale structure formation (\cite{Dod03}:162-173).
\end{itemize}
These effects all evolve in the forward direction of time that
underlies the expansion occurring then (Figures 6.7 and 6.8 in
\cite{Dod03}). It is this expansion and arrow of time that sets the
context for these important physical effects, being an example of
\textbf{AT3}. However the inflation would not start if the universe
at the beginning of the inflationary epoch was not of limited
anisotropy \cite{RotEll86} and inhomogeneity \cite{Pen89a}
(an example of \textbf{AT2}). \\

Inflation ends at reheating, when the inflaton gets converted to
radiation. This sets the almost homogeneous initial conditions for
the next stage.

\subsection{Epoch 2: The Hot Big Bang Era}\label{epoch2} In the
Hot Big Bang epoch, there was a heat bath with matter, photons and
neutrinos (\cite{Dod03}:40-46) mainly in equilibrium at a
temperature $T(t)$ that decreases with time. The early part of this
epoch is radiation dominated, but it becomes matter dominated before
last scattering (\cite{Dod03}:50-51). In the early radiation
dominated era, the scale factor goes as (\ref{frw2}) and the
temperature varies as (\cite{Dod03}:4-5)
\begin{equation}\label{Tradn}
T(t) = \frac{a(t_0)}{a(t)}T_0 \propto \frac{T_0}{(t-t_1)^{1/2}},
\end{equation}
decreasing as $t$ increases. This sets up a context which provides
the arrow of time for local reactions: the time symmetry is broken
by the steady drop in temperature of the heat bath that is the
time-changing environment
for local reactions such as nucleosynthesis.\\

The expansion of a equilibrium hot equilibrium mixture of particles
and radiation is time reversible (pair production and annihilation,
element formation and decomposition balance) until reaction
thresholds are passed so some of these reactions cease and leave
behind out of equilibrium decay products, characterizing
irreversible behavior at those times. The major such
non-equilibrium features are
\begin{itemize}
  \item Baryogenesis processes and the development of an
  asymmetry between particles and antiparticles (\cite{TurSch79}; \cite{Sil01}:134-137;
  \cite{Rio98});
  \item Neutrino decoupling, neutron decoupling, and the
formation of light elements at the time of nucleosynthesis
(\cite{Sil01}:140-143; \cite{Dod03}:9-12,62-70);
  \item Recombination of electrons and protons into neutral hydrogen,
resulting in decoupling of matter and radiation
(\cite{Sil01}:162-163; \cite{Dod03}:70-73), so determining the LSS
and originating the CBR (photons stream freely from then
on);\footnote{There is an earlier process of helium decoupling,
which is not as important thermodynamically.}
  \item Possibly, production of dark matter (\cite{Dod03}:73-78).
\end{itemize}
These irreversible processes can all be described by suitable
versions of the Boltzmann equation (\cite{Dod03}:59-62;84-113).  A
crucial feature is
\begin{itemize}
  \item Growth of perturbations as different comoving scales
  leave and re-enter the Hubble horizon (\cite{Dod03}:180-213),
  and baryon-acoustic oscillations take place
  (\cite{Dod03}:224-230),
  accompanied by diffusion on some scales (\cite{Dod03}:230-234)
  and radiative damping of shorter wavelengths (\cite{Sil01}:176-180).
\end{itemize}
Reversible and irreversible processes in this period have their time
arrow set by the expansion of universe which determines how context
changes: a time dependent heat bath where the temperature decreases
as a result of \textbf{AT2}. Initial conditions were special, which also plays a
role: the expansion rate and hence light element
production would be different in very anisotropic or inhomogeneous
cosmologies, so this is also an example of \textbf{AT2}. \\

Overall, this epoch serves to prepare special conditions on the LSS,
homogeneous to one part in $10^{-5}$, which marks the end of this
epoch when $T_{rad} \simeq 4000K$.

\subsection{Epoch 3: The astronomical arrow of time}\label{epoch3}

In this epoch, matter and radiation are initially decoupled, so they
are no longer in thermal equilibrium. Radiation pressure (which
previously led to the baryon acoustic oscillations) no longer
resists gravitational collapse, and structure formation can commence.\\

Structure formation takes place spontaneously through gravitational
instability (\cite{Sas87}: Chapter 21). An initially uncorrelated
system develops correlations through gravitational attraction:
``\emph{Gravitational graininess initiates clustering}''
(\cite{Sas87}:158-162). There is no arrow of time in the underlying
time-symmetric Newtonian gravitational law:
\begin{equation}\label{newt}
m \frac{d^2x^i}{dt^2} = - \nabla^i \Phi,\,\, \nabla^2\Phi = 4\pi G
\rho
\end{equation}
(which derives in the appropriate limit from the Einstein Field
Equations). The process attains an arrow of time in the expanding
universe context because of the change of equation of state at the
LSS: pressure forces that resisted collapse melt away, and structure
formation begins from the tiny density inhomogeneities present on
the LSS. The arrow of time is then provided by the context of cosmic
evolution, communicated from the global scale to local scales by
passing down the expansion parameter $H(t)$ which occurs in the perturbation equations \cite{EllBru89}.\\

Structure forms spontaneously through a bottom up process
dominated by cold dark matter (\cite{Ree95}:23-35,39-45;
\cite{Sil01}183-186); this process apparently violates the second
law of thermodynamics as often stated (\cite{Pen89,Ell95};
\cite{Car10}:295-299); see Figure 1d. The entropy law somehow must
be consistent, but we do not at present have a viable definition of
gravitational entropy for such situations. In any case it is clear
that the process requires special smooth initial conditions so that
structure can form: if black holes were already present everywhere,
there would be no possibility of further structure formation
\cite{Pen89}.\\

Star formation takes place leading to ignition of nuclear fusion and
 nucleosynthesis in stellar interiors
(\cite{Sil01}:301-331;339-345). The CBR is now
collision free (\cite{Sil01}:75-81), so its temperature drops as the inverse of the scale
factor (\cite{Dod03}:5):
\begin{equation}\label{Tradn1}
T_{radn}(t) = \frac{a_{LSS}}{a(t)}\,4000 \,K.
\end{equation}
This is much lower than the stellar temperatures, so the stars can
function in thermodynamic terms (they can get rid of heat by
radiation to the sky). The many irreversible astrophysical processes
\cite{Ree95,Shk78} leading to the evolution of stars
(\cite{Sil01}:202-204, 229-231, 288-298) and galaxies and galaxy
clusters (\cite{Sil01}:187-202, 323-327) are based in bottom-up
emergence of effective thermodynamical behavior, but this is
possible only because of the non-equilibrium context set by the
early universe according to the cosmological master arrow of time.
The lowness of the Sun's entropy (remoteness from thermal
equilibrium) is because of the uniformity of the gas from which the
Sun has gravitationally condensed (\cite{Pen04}:705-707).\\

\textbf{Conclusion}: Structure formation takes place by irreversible
processes starting in the inflationary era, resulting in
fluctuations on the LSS that irreversibly lead to stars, galaxies,
and planets after decoupling. The arrow of time for these processes
derives from the cosmological master arrow of time (Section
\ref{epoch2}).

\subsection{Thermodynamic arrow of time: local
systems}\label{epoch4} For local systems on Earth, the arrow of
time is apparent in the diffusion equation and in local physical
interactions in machines, plants, animals, ecosystems, and the
biosphere as a whole (\cite{Zeh07}:39-84). This is all possible
because we live in a non-equilibrium local environment, which is
due to the larger astronomical environment
(Section \ref{epoch3}).\\

\textbf{Bright Sun plus dark night sky} The Sun is a radiation
source which is a hot spot in an otherwise cold background sky.
Because of their higher energy, there are many fewer photons coming
in from the Sun than those reradiated in the infrared to the sky,
since the total energy carried in is the same as that going out
(\cite{Pen89}:415; \cite{Pen04}:705-707; \cite{Car10}:191-194). The
radiation heat balance equation for received solar short wave
radiation (\cite{Mon73}:60-61) leads to overall annual, daily and
instantaneous heat balances (\cite{Mon73}:71-77) due to the
properties of incoming solar radiation (\cite{Mon73}:23-58) and
radiative properties of natural materials (\cite{Mon73}:60-71). This
leads to the heat balance equations for animals that enables life to
function (\cite{Mon73}:150-170). This is all possible because the
sky acts as a heat sink for the emitted long wavelength radiation.\\

The reason the sky can act as a heat sink is the modern version of
\textbf{Olber's paradox} (why the sky is dark at night?
\cite{Har87}, \cite{Har00}:248-265). The sky is dark because the
universe is expanding (\cite{Har00}:491-506; \cite{Sil01}:55-58), so
by (\ref{Tradn1}), the Cosmic Background Radiation has cooled from
its temperature of 4000K at the LSS to the present day CBR
background temperature of 2.75 (\cite{Har00}:339-349;
\cite{Zeh07}:26-27). Star formation since decoupling has made a
negligible further contribution: it has led only to an effective
temperature of 3K also. This would not be the case if there were a
forest of stars covering the whole sky (\cite{Har00}: Figure 12.1,
p.250), when every line of sight would intersect a star and the
temperature on Earth would be the same as on the surface of a star.
Thus we are in a thermal bath at $3K$ as a result of expansion of
universe and its subsequent thermal history, resulting both in
cooling of the CBR to 2.73K and in stars only covering a very small
fraction of the sky. This astronomical context underlies the local
thermodynamic
arrow of time.\\

\textbf{Example: Broken glass} A classic example is a glass falling
from a table and lying shattered on the floor (Penrose
\cite{Pen89}:397-399). Because the underlying micro-dynamics is
time-reversible, in principle it can be put together again by just
reversing the direction of motion of all the molecules of the glass
and in the air and the floor: it should then jump back onto the
table and reconstitute itself. But this never spontaneously happens.
Why is the one a natural event and the other
not?\\

It does not spontaneously reverse and reconstitute itself
because this is fantastically improbable: the asymmetric increase of
entropy, due to coarse graining, prevents this
(\cite{Pen89}:391-449); and that is because of special conditions
with correlations (the crystal structure) in the initial state that
don't occur in the final state.

\subsection{Isolated systems and the Radiative arrow of time}\label{radiate}
There is a further important issue: the relation between the
radiative arrow of time and the thermodynamic arrow of time
\cite{EllSci72}. Consider first water waves spreading out,
consequent on a stone being thrown into a pond. In principle,
because of the time reversible microphysics, one can reverse the
direction of time to see the waves focus in and make the stone pop
out of water. In practice this can't be done. Again, we have a
resolution by asymmetric correlations: typical incoming
waves are not correlated, but the outgoing waves are (they diverge
from one point). Thus the arrow of time is reflected in the
asymmetry of correlations in the future relative to the past. \\

But is this asymmetry a cause or an effect? Suppose we don't want to
talk about the future: can we just talk about special initial
conditions in the past?  Yes, this should be possible: all we need
is the structure of phase space (\cite{Pen89}:402-408) plus special
conditions in the past (\cite{Pen89}:415-447). One does not need a
future condition. But one does need the past condition \emph{on the
relevant scale} (it is needed on the scale of stars for the
astrophysical arrow of time, but on the scale of molecules for the
arrow of time in water waves). Hence we need a
\begin{quote}
\textbf{Local Past Condition (LPC)}: \emph{special initial data
occurred at the relevant scale for the phenomenon considered}.
\end{quote}
 The
initial conditions that lead to structure are less likely than those
that don't, but they did indeed occur in the past. This \textbf{LPC}
applies at the scale relevant to broken glasses and unscrambling
eggs because it has cascaded down from the cosmological scale to the
local scale. This incoming and outgoing asymmetry applies to water
waves, sound waves in the air,  and elastic waves in solids.  It
applies on the astronomical scale to supernova explosions: one can
in principle reverse the direction of time to see the outgoing radiation
focus in and the supernova reassemble; in practice this cannot
happen because of the very special initial conditions required for
the time reverse motions to do this.

\subsubsection{Electromagnetic waves} The same issue arises for
electromagnetic radiation, indeed this is \emph{the} relevant case
as regards the heat from the Sun because that arrives as radiant
energy. Why does it come from the past null cone rather than the
future null cone, given that Maxwell's theory is time symmetric? And
why do radio signals arrive after they are sent, rather than before?
The answer is similar to that for acoustic waves: there is a
fundamental difference between incoming and outgoing electromagnetic
radiation, in terms of coherence on the future null cone as compared
with the past. But then why is that so? It derives from cosmic
initial conditions, cascading down from larger to smaller scales. To
look at this properly, we need to be clearer on the spacetime
domains involved.

\subsubsection{Isolated systems: The relevant domains}\label{isolated} We need to
consider an effectively
 isolated system, such as the Solar System (see Figure 3, which is in conformal coordinates, with matter world lines
vertical lines). \emph{The Earth's world line} is at the center, the
event\emph{`here and now'} is where the present time intersects our
world line. The incoming light cone (to the past) intersects the LSS
on a 2-sphere, which I call the \emph{CMB sphere}, because this is
the part of the LSS that emitted the radiation we today measure as
2.73K Cosmic Black Body Radiation (the CMB sphere has been mapped in
detail by the WMAP and Planck satellites). The \emph{Visual Horizon}
is formed by the matter world lines through the CMB sphere (we
cannot see any matter further out by any electromagnetic radiation).
Hence the whole visible universe lies between our world line and the
visual horizon, from the LSS till today.
On these
scales we can extrapolate to the future, but with increasing
uncertainty
the further we extrapolate towards the final future events.\\

To examine the relation between incoming and outgoing radiation, one
can use the idea of \textbf{Finite Infinity}  ${\cal F}i$
\cite{Ell84}. We surround the system $S$ of interest by a 2-sphere
${\cal F}i$ of radius $R_{{\cal F}i}$ such that it is at infinity
for all practical purposes: spacetime is almost flat there, because
it is so far away from the source at the center, but it is not so far
out that the gravitational field of other neighboring objects is
significant. For the Solar System, $R_{f}$ is about 1 light year;
for the Galaxy, 1 Mpc. The world tube marked out by ${\cal F}i$ is
shown in Figure 3; we can examine the interaction of the local
system with the rest of the universe by considering
incoming and outgoing matter and radiation crossing this world tube.\\

The intersections of our past light cone $C^-$ with ${\cal F}i$
gives a 2-sphere  ${\cal C}^-$ a distance $R_{{\cal F}i}$ away which
is our effective sky; all incoming radiation crosses ${\cal C}^-$.
Similarly the 2-sphere ${\cal C}^+$ defined by the intersection of
our future light cone with ${\cal F}i$ is our future sky; all
outgoing radiation crosses ${\cal C}^+$. The arrow of time has two
aspects. First, it lies in the difference between data on ${\cal
C}^+$, which high correlations with our position due to outgoing
signals from the Earth, whereas that on ${\cal C}^-$ does not have
time-reversed similar correlated incoming signals focussed on the
Earth. Second, it lies in the fact that the amount of incoming
radiation on ${\cal C}^-$ is very low; this is the dark night sky
condition mentioned in the previous section. It is in effect the
Sommerfeld incoming radiation condition (\cite{Zeh07}:23).\\

The reason there is little radiation coming in on ${\cal C}^-$, and
that it is uncorrelated to a high degree, is two-fold. Firstly,
there is a contribution from the radiation temperature on the CMB sphere on the
LSS, which is almost Gaussian, and then is diluted by the cosmic
expansion from 4000K to 2.75 K (see above). Second, the intervening
matter between the LSS ${\cal C}^-$ and us is almost isotropic when
averaged on a large enough scale, and luminous matter covers a
rather small fraction of the sky (we do not see a forest of stars
densely covering the whole sky); hence we receive rather little
light from all this clustered matter (stars in our galaxy, all other galaxies,
QSOs, etc). \\

But how does this relate to solutions of Maxwell's equations in
terms of advanced and retarded Green's functions
(\cite{Zeh07}:16-38)? And why does matter here, and the intervening
matter, emit radiation to the future rather than the past? This is
allowed by thermodynamic constraints on the emission processes; but
this does not by itself explain the electrodynamic arrow of time.
Why does a shaken electron radiate into the future and not the past?
We need a condition where the waves generated by a source are only
waves that go outward, so only the outgoing wave solution makes
physical sense (\cite{FeyLeiSan64}:20-14). I suggest that the reason
is that only the past Green's function can be used in such
calculations, because we live in an Evolving Block Universe (EBU):
we can't integrate a Green's function over a future domain that does
not yet exist. This is discussed in Section \ref{sec:time1}. The
Local Past condition \textbf{LPC} is needed on these scales so that
thermodynamic and associated electrodynamic processes can take place
in the forward direction of time; but the foundational cosmological
arrow of time \textbf{AT3} in the EBU is the ultimate reason that times goes to
the future and not the past. Actually it defines what is the future
direction of time.

\subsubsection{Incoming and outgoing matter} As well as radiation, an
isolated system is subject to incoming and outgoing matter. There
are two aspects here.  First, the matter that made up the solar
system and nearby other systems -- indeed the matter out of which we
are made
--- originated within our \textbf{matter horizon} \cite{EllSto09}, a
sphere with comoving radius of about 2 Mpc, lying between ${\cal
F}i$ and the visual horizon (Figure 3). The intersection of the matter horizon
with the LSS (see \cite{EllSto09} for a discussion and detailed
spacetime diagram) is the domain where we require special past
conditions to be true in order that the solar system can arise from
astrophysical processes with the forward
arrow of time. This is the \textbf{LPC} for the Solar system.\\

Second, there may be incoming particles (cosmic rays, black holes,
asteroids, comets) crossing ${\cal F}i$ since the solar system
formed and impacting life on Earth. These too must be of low
intensity in order that local equilibrium can be established and
local thermodynamic processes proceed unhindered. The Local Past
Conditions on the LSS in the domain close to the matter horizon will
ensure this to be true. We assume this for example in experiments at
particle colliders such as the LHC: if there was a huge flux of
incoming cosmic rays, we would not be able to do experiments such as
at the LHC. Local thermodynamics can proceed as usual because we are
indeed an effectively isolated system. The universe does not interfere with our local affairs: a case of
\emph{top-down non-interference} that could have been otherwise:
there could have been massive gravitational waves or streams of
black holes coming in and interfering with local conditions, as well
as high energy photons. Isolated systems are necessary for life
(\cite{Ell06a}: Section 9.1.3), and the cosmic context sets this
local context up suitably.

\subsection{Micro systems: quantum arrow of
time}\label{epoch_micro} The same kinds of considerations hold for
everyday physics and local quantum systems. They work in the
forward direction of time because of the non-equilibrium local
context inherited from the higher level solar system context. If
we were in a higher temperature heat bath,
there would be different outcomes.\\

The quantum arrow of time (\cite{Zeh07}:85-134) should follow from
the local context, because for example the way wave function
collapse occurs in detectors (based in the photoelectric effect) is
due to the local physical context \cite{Ell11a}. That context
includes the local thermodynamic arrow of time. So for example one
can ask why photodiodes or chlorophyll in plant leaves don't behave
reversibly: why does a plant or a CCD not emit light rather than
absorbing it? The answer must be that they involve special
structures that create thresholds that general non-unitary
behavior, and the specific arrow of time that occurs is set by
prepared initial conditions in the physical apparatus, that make the
detector work in the one direction of time, not the other, in the local context discussed above. This is
related to the general feature of anisotropic spatial structures
plus special initial conditions.\\

 The derived quantum arrow of time is
synchronized with the overall system through contextual effects, and
in particular decoherence and the Lindblad master equation inherit
their arrow of time from the environment. Each is not time
reversible because the environment is in a non-equilibrium state
(Section \ref{time4}) and a local asymmetry condition \textbf{LPC}
applies on micro scales: as in the case of reconstructing a
supernova, one could not easily reverse the chemical reactions in
photosynthesis, as this would require improbable coordination of
incoming entities.

\section{The ascent of time: emergent structures}\label{time6}
The existence of time, and the direction of
the arrow of time, is taken for granted in applied physics,
engineering, biology, geology, and astrophysics. It is assumed as
a ground rule  that the arrow of time exists and runs
unceasingly according to the 2nd law. Given the arrow of time problem as set out above, how does this come about?\\

The suggestion here will be that the arrow of time that exists at
the lower levels (because of the suitable context, as discussed in
the previous section) propagates up to higher levels through the
process of creation of emergent structures. This is a bottom up
process from lower to higher levels.
There are specific emergent mechanisms that enable this to happen,
with three related crucial components: arrow of time detectors
(Section \ref{detect}), rate of time measurers (Section
\ref{clocks}), and flow of time recorders (Section \ref{records}).
These are what make the flow of time real. I look at them in turn.

\subsection{Arrow of time detectors}\label{detect}
 If we take series of
pictures of objects such as a breaking glass or an exploding
supernova,  we can discriminate the future from the past by just
looking at them: we can order the pictures appropriately and
determine the arrow of time. But these aren't regular occurrences
that can be used generically to determine the direction of time;
there are more systematic ways of doing this. \\

Generically, a cause precedes its effects; how does one harness this
to show which way time is going? The basic principle is
\begin{quote}
\emph{\textbf{A spatial asymmetry  is converted into a time
asymmetry through suitable environmental and initial conditions.}}
The requirements are structures with suitable spatial asymmetry,
plus special initial conditions.
\end{quote}
This is a form of top-down action, due both to the existence of
emergent structures, and the link in to the \textbf{LPC} discussed
above through the requirement of special initial conditions.
Specific cases show how this works out in detail.\\

-----------------------------------------------------------------------------------\\

\begin{figure}
\begin{center}
\label{FIGURE 4}
\includegraphics[width=6.0in]{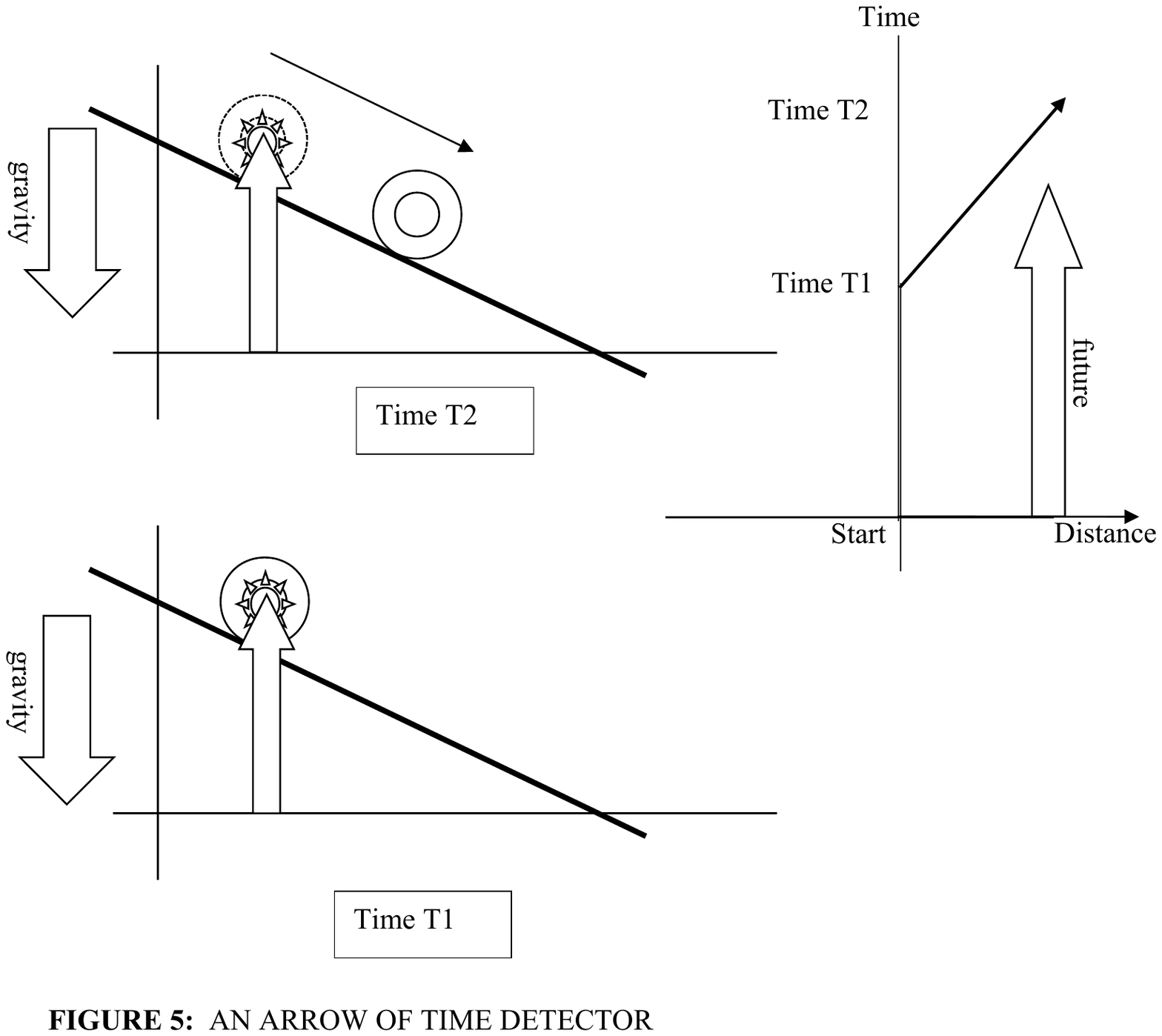}
\end{center}
\end{figure}

FIGURE 5 HERE\\

Caption: AN ARROW OF TIME DETECTOR\\ 

\texttt{{\small A wheel is clamped in a static
state for an extended period on a downhill slope; during that time
dynamics does not distinguish an arrow of time (because there is no
dynamics!). The slope establishes a spatial asymmetry.
At time $T_1$
it is released. The time $T_2$ when it has rolled some distance down
the slope is later than time $T_1$: the rolling down the slope (to
the right) establishes which is the past and which the future
direction of time.
If the wheel had not been clamped before the
start instant, the motion would have been time symmetric (it could
have rolled up to an instantaneous stationary state at time $T_1$
and then down again) and no arrow of time would have been
determined by the dynamics. } 
}\\

-----------------------------------------------------------------------------------

\begin{itemize}
\item \textbf{Downhill flow} A rock spontaneously falls down hill, not up; water naturally flows
downstream, not up (Figure 4). This applies to \emph{any} energy
gradient: exothermic reactions take place spontaneously in chemistry
(hence the danger of fires and the need for fire prevention
services); electric currents flow from the negative to the positive
terminal of a battery, and in electrical and electronic circuits,
currents flow towards ground. A natural example is lightning
 which goes spontaneously to the
ground (\cite{FeyLeiSan63}:9-2 to 9-7). In electronics, forms of
electric current include the flow of electrons through resistors or
through the vacuum in a vacuum tube, the flow of ions inside a
battery or a neuron, and the flow of holes within a semiconductor; they each flow one way
as time progresses. A reversed arrow of time would reverse their spatial direction of movement.

Biology is crucially based in the absorption of high energy
materials and excretion of low energy waste, the central feature
being cellular respiration, based in glycolosis, the citric acid
cycle, and oxidation (\cite{CamRee05}:160-178). Each of these
processes has a forward direction of time that arises out of the
underlying physics plus the local context, and so acts as an arrow
of time detector.
\item \textbf{Ratchets} A mechanical ratchet turns one way because of a pawl and ratchet
mechanism permitting motion in one direction only as time progresses
(Figure 5), thus it is a mechanism for detecting the direction of
time as contained in Newton's law of motion. How does it do it? ---
it is a mechanism designed to do so! There is no direction of time
in Newton's law of motion itself, but there is in the special
solution of the law of motion that describes the ratchet: a case of
spontaneous symmetry breaking, converting spatial asymmetry in to
time asymmetry. It can only do so when suitable environmental
conditions are satisfied: in general Brownian motion takes place,
and the pawl can jump out of ratchet allowing it to fluctuate back:
when the pawl and are wheel both at the same temperature, the motion
is reversible (\cite{FeyLeiSan63}:46-1 to 46-9). At lower
temperatures it is a disguised form of asymmetric sawtooth
potential, where diffusion extracts the direction of time from the
spatial direction provided by the sawtooth.

Engineering applications include ratchet wrenches and screwdrivers,
turnstiles, and hoists. Ratchets are a key mechanism in
microbiology: Ref. \cite{Seretal07} describes a molecular
information ratchet, \cite{Roe11} organic electronic ratchets doing
work, and \cite{Kul11} botanical ratchets. Brownian ratchets work by
inscribing and erasing an asymmetric potential which induces a
directed motion of a particle. Molecular motors are based on
biological ratchets \cite{LacMal10}, and work by hydrolizing ATP
along a polar filament.

 \item \textbf{Rectifiers} are devices that make currents flow in only one
 direction. If you reverse the direction of time, it will go the other
 way. This works thermodynamically in the case of a vacuum tube rectifier  (emission of
  electrons at a hot electrode and reception at a cooler one). It works by adaptive selection through
  detailed physical structure of the rectifier, in the case of solid-state
  rectifiers, taking advantage diffusion currents. In a p-n junction with forward bias, the
  electrostatic potential in the n-region is lowered relative to p-side, increasing the diffusion current;
  the pair current is unchanged. The diffusion current exceeds the pair current
  and there is a net current from the p-side to the n-side; however
  with a reverse bias this does not happen, hence the junction acts as a rectifier (\cite{Dur00}:997).
A mechanical example is a one-way valve in a water system, with a ball and spring in a water outlet into a container.
Water flows only in through this valve, not out.\\

A crucial example of rectifiers in biology is ion channels
(\cite{KanSchJes00}:105-124; \cite{RhoPfl96}:219-226;
\cite{CamRee05}:133-136, 1017-1025). These are pore-forming
proteins that establish and control a voltage gradient across the
plasma membrane of cells, thereby allowing the one-way flow of ions
down their electrochemical gradient. They occur in the membranes
that surround all biological cells, for example potassium ion
channels (the hERG channel) mediates a delayed rectifier current
(IKr) that conducts potassium (K+) ions out of the muscle cells of
the heart \cite{Truetal95}. All such
rectifier functions are based in detailed biological mechanisms
(e.g. \cite{Goeetal02}). These underlie active transport systems in
the cell that act like Maxwell's Demon in creating a gradient of
$K^+$ and $Na^+$ across a cell wall (\cite{Leh73}:191-206).

 \item \textbf{Filters}
 are devices that select some components of a mixture or ensemble from others, discarding those not
 selected. The arrow of time is revealed by the process of selection
 where a subset of the whole emerges as the output (Section \ref{adapt}). Running it in
 reverse would generate more states from less, but the final state does not
 have the information needed to tell what the incoming state was. Examples are
 polarizers \cite{Ell11a}, wavelength filters in optics due to selective absorbtion
 based in the crystal structure of an optical medium, and tunable radio receivers based in resonance properties
 of AC circuits (\cite{Hug08}:299-320). Electrical filters can be low-pass, high-pass, or passband filters
 (\cite{Hug08}:359-385). In biology, excretory processes in
 organisms are based on a series of filter mechanisms involving
 selectively permeable membranes and selective reabsorption
 (\cite{CamRee05}:929).

  \item  \textbf{Diffusion} is a basic physical process that detects the direction of time:
  macro-properties of gases naturally smooth out in the future, not the past (Figure 1).
  Thus diffusion is migration of matter down a concentration gradient (\cite{Atk94}:818).
  Similarly migration of energy down a temperature gradient underlies thermal conduction,
  migration of electrical charge down a potential gradient underlies electrical conduction,
  and migration of linear momentum down a velocity gradient generates viscosity
  \cite{Atk94}:818). This is a crucial process in chemistry
  (\cite{Atk94}:817-830,846-856) and thermal physics
 (\cite{Fuc96}:649-654).

Physiological processes often involve diffusion between compartments
and through membranes, the direction of time determining the
direction of diffusion as an emergent property of the lower level
dynamics (\cite{Rig63}:168-220, \cite{RhoPfl96}:116-129). Diffusion
plays a key role in capillary systems (\cite{RhoPfl96}:590-592),
hormone transport (\cite{RhoPfl96}:373-374), and lungs
(\cite{RhoPfl96}:618-619), where it determines action of an
anesthetic gas (\cite{Rig63}:312-333). Diffusion is crucial at the
synapses connecting neurons in the brain (\cite{CamRee05}:139-146,
\cite{RhoPfl96}:113-116).

 \end{itemize}

\noindent In each case, the macro context acts down on the micro
level to induce time asymmetric behavior arising from spatial
gradients (either horizontal, so unaffected by gravity, or with a
vertical component, so gravitationally influenced). The macro level
dynamics due to inhomogeneities at that level acts down on the micro
level to create micro differences (the temperature of a falling rock
is hotter at bottom of hill than at top).\\ 

-----------------------------------------------------------------------------------\\

\begin{figure}
\begin{center}
\label{FIGURE 5:}
 \includegraphics[width=6.0in]{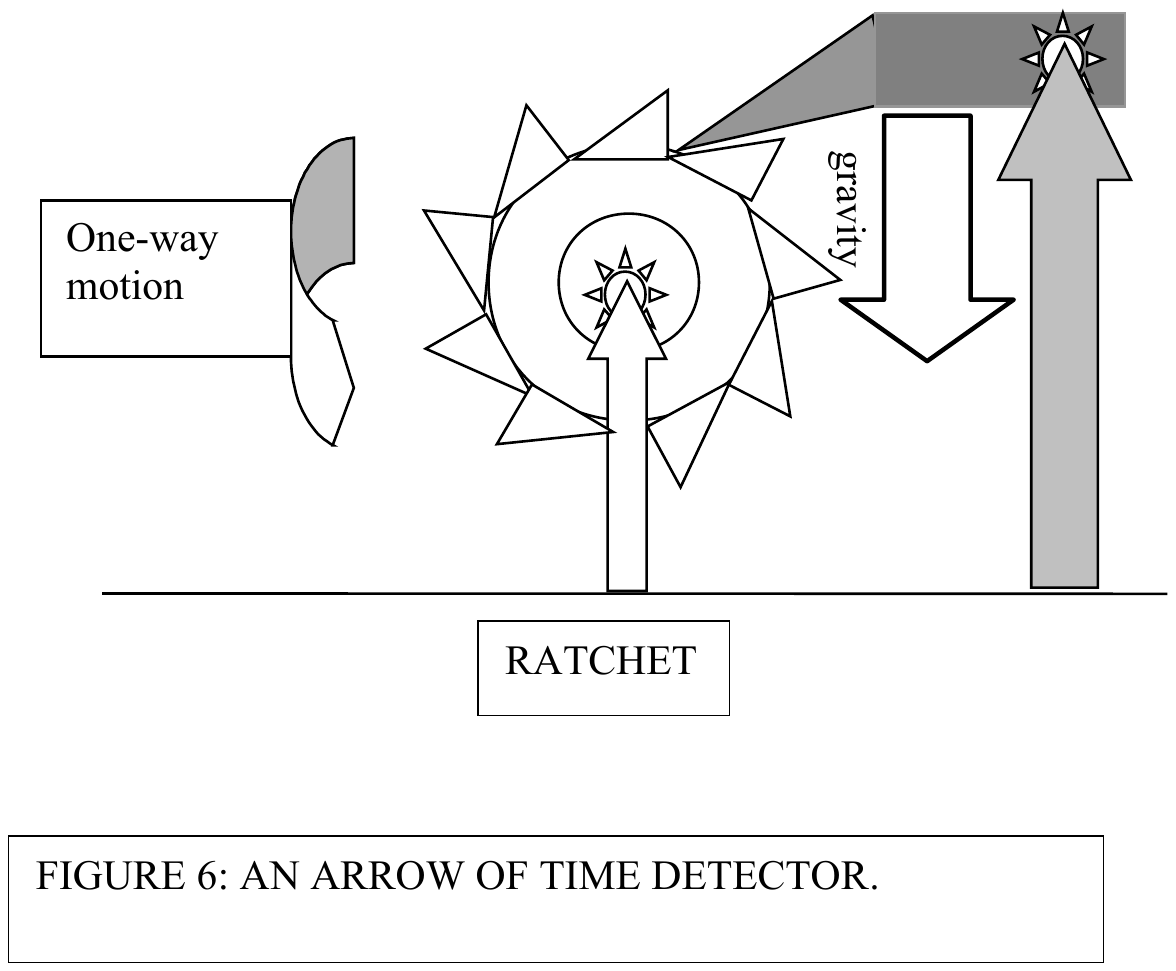}
\end{center}
\end{figure}

FIGURE 6 HERE\\

Caption: AN ARROW OF TIME DETECTOR:\\ 

\texttt{{\small A ratchet wheel is constrained
to rotate only one way by a pawl held down by a spring or by
gravity. Its one way motion characterizes the future time occurring
in the Newtonian equations of motion.  Its
environment must be cool. } 
}\\

-----------------------------------------------------------------------------------


\subsection{Rate of time measurers: clocks and ages}\label{clocks}
 \textbf{Clocks} are rate of instruments that reliably measure the rate of progress of time, converting it to a
linear scale; an integrated clock reading gives an age. Translating it to digital form
will always involve non-unitary events.\\

Some clocks are direction of time detectors in addition (any of the
arrow of time detection processes mentioned above can be used as a
clock, if it's behavior is regular enough). Many clocks do not
measure the direction of time but merely the number of intervals
between two events, independently of whether those intervals run
from $a :\rightarrow b$ or are reversed to run from $b : \rightarrow
a$. These kinds of clocks are represented by solutions which do
exhibit the time-symmetry of the underlying laws of motion. A light
clock is one such example, a ratchet clock not.


\subsubsection{Reversible clocks}
Clocks that are not direction of time detectors are as follows.\\

\textbf{Distance traveled at reliable speed} measures time. An example is an analog
clock that rotates its hands at a constant speed through an electric
motor drive. The non-unitary part of the process occurs when a dial
reading is noted at a specific time by some observer. A sundial is a
projection to Earth of such a reliable motion (albeit non-constant:
seasonally dependent but predictable corrections must be made). Some
forms of clock use the invariance of the speed of light to provide a
fundamental basis for timekeeping. A simple example is a \emph{light
clock} consisting of an emitter and two mirrors kept a fixed
distance apart; the `ticks' of the clock are the reflection events
at one end. The non-unitary part of the process is the reflection
events.\\

\textbf{Counter for repetitive processes} Most current clocks
work by counting cycles of some reliable periodic process, like the
swing of a pendulum, the cycles of a balance wheel, or the vibration
of a quartz crystal. A reader or latch records the cycle, and this
analog to digital transformation involves thresholds and so is
non-unitary. A computer utilizes a circuit that emits a series of
pulses with a precise width and a precise interval between
consecutive pulses, made by an oscillator and latch (a circuit that
remembers previous values)  (\cite{Tan90}:98-103). This is of course
a classical device; with an adder (\cite{Tan90}:96-98), it detects
the direction of time and measures the progression of time. It must
be based in repeated wave function projection at the quantum level
in the transistor gates that underlie its operation
(\cite{Tan90}:76-86).

\subsubsection{Irreversible clocks}
Clocks that are direction of time detectors are as follows.\\

\noindent \textbf{Dating} One can estimate ages of objects by
examining historical records, for example in astronomy, archaeology,
and geology. Accurate age measurements come from integrating clock readings.\\

\textbf{Reliable decay processes}  Radioactive materials
spontaneously decay; the future direction of time is that choice
where the amount of radioactivity is less in the future than the
past (\cite{FeyLeiSan63}:5-3 to 5-5). This provides centrally
important radiocarbon methods of dating in archaeology
(\cite{Woo05}:31-33). This is a form
of the emptying reservoir method mentioned above; state preparation
took place in the supernova explosion that created the radioactive
elements.\\

\textbf{Flow in and out of a container} One of the oldest methods
  is a container with a steady inflow and/or outflow of some quantity, and suitable
  calibration. It requires a prepared initial state (full or empty)
  and identifiable final state, when it will require a reset to the initial
  condition; this is the non-unitary part of the process. Examples
  are water clocks (filled to the brim and then steadily emptying) and sand
  funnels (egg timers). One can conversely have a water container that is
  steadily filled up, and then flushed when full. In electrical circuits, charging and
  discharging a capacitor is the equivalent (\cite{Hug08}:93-94;108-114).\\

\textbf{Reliable growth processes} can also be used for dating. The classic case
(apart from horse's teeth!) is dendochronology: that is dating by
counting tree rings (\cite{Woo05}:34). Actually this is really a
process of recording the annual cycles of the Earth's motion around
the Sun using the tree's developmental processes as the recorder. In
astronomy, stellar ages can be determined via stellar evolution
theory and observations of the distribution in the
Hertzsprung-Russell diagram of cluster stars (\cite{Shk78}:189-193).\\

All of these measuring processes result as emergent properties of
the underlying physics that at some point involves non-unitary
evolution (without this feature, the flow of time would be evident
in the system dynamics but it could not be recorded).

\subsection{Flow of time recorders: records of the past and
memory}\label{records}
 We are aware of the flow of time because of the existence
of records of the  past. These are of two types:
\begin{quote}
\textbf{Passive records}: these are physical records of what
happened in the past, such as primordial element abundances,
geological strata, palaeomagnetic records, fossils, the genetic
code, the nature of biological species, vegetation patterns in the
countryside, buildings and infrastructure in cities, and so on. This
is \emph{data} which can be used to provide information about the
past, when we relate them to some theoretical model.
\end{quote}
These data have not been laid down for some purpose; they are just
there as remnants of past events. They have not been indexed or
classified, but we are able to interrogate them and use them to
determine past history, for example in astronomy \cite{Shk78,Sil01},
geology, and biological evolution (\cite{Woo05}:24-35).
\begin{quote}
\textbf{Memory records}: these are physical records that are a
meaningful distillation of what happened, somehow indexed in
relation to some classification scheme, so that they can be
recovered. This is \emph{information} that can be used for some
useful purpose \cite{Roe05}.
\end{quote}
In both cases, laying down these records of events that have
happened involves a physical process with a definite physical
outcome which is then stable over some length of time.

\subsubsection{Recording} Some physical property is set at a specific value
by a local physical process, and then stays at the value because a
threshold has to be surmounted to reset it. The arrow of time comes in that the record did not initially exist but does at later times.\\

The kinds of properties that serve as the physical substrate at the
lower levels include:
\begin{itemize}
\item \emph{\textbf{Magnetic states of a magnetizable medium}}, based in the magnetization properties
of specific materials (\cite{FeyLeiSan64}:36-1 to 37-5).
\item \emph{\textbf{Coded surface properties of a medium}}: Binary data is stored in pits on the surface
of CD ROMs (\cite{Tan90}:52-54)), based in the stability of material
properties. One can also include in this category, writing and
printing on paper.
\item \emph{\textbf{Electric circuit states}}: Binary data is stored in
high/low voltages in specific circuit elements in some memory array,
based in the stability of electronic circuit states
(\cite{Hug08}:544-554).
\item \emph{\textbf{Biomolecule structure}}: the coding patterns in
DNA molecules is a record of evolutionary history, and can be used
for dating and cladistic analysis (\cite{Woo05}:21,51-52). This is
based in the reliability of the DNA copying process
(\cite{CamRee05}:293-308).
\item \emph{\textbf{Pattern of connectivity and activation in network connections}}:
The specific pattern of connections and their relative strengths in
neural networks records short term and long term memories
\cite{KanSchJes00}:1227-1246). Short term data can also be stored in
as activation patterns such as
 synchronized activity in brain circuits (\cite{Buz06}:136-174).
\end{itemize}
Generically, recording takes place through emergence of any
identifiable structure that
is stable over a relevant time scale.

\subsubsection{Remembering} This is some process whereby the stored record is
interrogated: examples are DNA epigenetic processes, reading of
memory in computer systems, and recalling memories in one's mind.
This is as opposed to interrogating a record, where the record is
analyzed rather than being read, as in the cases of geology and
archaeology. For memory to be useful, there has to be some kind of
indexing system of what is stored: it is no good storing
information, if you don't know where it is.

The process of indexing needs a sorting and classification system,
which will generally involve a modular hierarchical structure used
in the classification of folders and files and in assigning names or
other identifiers to them as well as to stored entities. In a
computer it is implemented at the lower level by association of
names with specific memory addresses (\cite{Tan90}:40-44). The index
is itself a further form of memory.

\subsubsection{Deleting}
Because of the finite capacity of memory and the
ongoing influx of new information, generically one needs some kind
of reset process that wipes out old memory to create space for new
information to be stored. This is state preparation for the next
round of remembering. It will be a non-unitary process, indeed it is
precisely here that irreversibility occurs and entropy is generated,
as shown by Landauer \cite{Lan61} and Bennett\cite{Ben03}.
The arrow of time comes in that the record that initially existed does not exist at later times.\\

 However you don't delete all
that is in memory: \emph{selective deletion} takes place, because
one selects what is deleted and what is kept by deleting unwanted
files, emails, and so on. this is therefore a form of adaptive
selection: the creating of useful information deleting that which is
not useful in relation to some classification system and guiding
purpose. What is kept is determined by the user's purpose: this is
top-down causation from the user's purpose to the electrons in the
computer memory system. The arrow of time is involved in the transformation of random records to useful data.

\subsubsection{State vector reduction}
The crucial feature of all of this in relation to quantum physics is
that \emph{\textbf{every recording event, reading event, and
deletion event involves effective collapse of the wavefunction}},
because a recallable classical record is laid down and has a
definite state. Quantum uncertainty makes way to classical
definiteness as each such event takes place. This is happening all
the time everywhere as passive records are laid down, as well as
when memories are recorded, read, and deleted. The outcome is no
longer a quantum superposition: it is a definite classical outcome.
If this was not the case, specific memories could not be recalled.

\subsection{The ascent of time: emergent
properties}\label{time61} The proposal now is that the passage of
time, which happens in events such as those just discussed, ripples
up from the lower levels to the higher levels through developmental
processes that depend on the lower level arrow of time and therefore
embody them in higher level processes.\\

The process of emergence builds an arrow of time into each higher
emergent level $N+1$ because it is imbedded in the next lower level
$N$, through the process whereby coherent lower level dynamics leads
to emergence of coherent higher level dynamics (Diagram 1). If you
reverse the arrow of time at the lower level in Diagram 1, it will
result in a reversed arrow of time in the higher level. For example,
if the future direction of time is built into machine language level
in a computer, so a program at that level runs in the positive
direction of time $t$ when the computer is run, the same will be
true at all the emergent abstract machine levels in the computer
\cite{Tan90}; for example a Java virtual machine running
on top of the machine language will have the same arrow of time.\\

The basic ways the lower level processes embody the arrow of time
has been discussed above: they include,
\begin{itemize}
  \item Things naturally fall downhill,
  \item Electric currents flow from positive to negative potentials,
  \item Waves convey information as they spread out from their
  sources,
  \item Energy changes from useful to useless forms as dissipative
  processes take place,
  \item Diffusion spreads heat and matter out from their sources.
\end{itemize}
These each embody an arrow of time: for example given a definition
of positive (+) and negative (-) potentials, the current flows from
the + to the - terminal \emph{in the future direction of time}. If
the direction of time were reversed, the flow would be the other
way.\\

These effects at the basic physics level then affect processes in
applied physics, chemistry, all forms of engineering, geology,
planetary science, and astronomy, as well as in microbiology,
physiology, developmental processes, psychological processes, and
evolutionary history, leading to similar effects due to the flow of
time. Indeed \emph{it is because this happens reliably that we have
our basic concepts of cause and effect}, the latter always occurring
after the former. The whole idea of causation is premised on this
property.

\subsubsection{Local physics and technology} Physicists, chemists,
and engineers can assume the 2nd law of thermodynamics holds on
macro scales with the forward direction of time. This becomes a
basic feature of their analyses \cite{Fuc96}, involving viscosity,
the production of heat, dissipation, and entropy production. It
affects entities such as engines, refrigerators, heat exchangers, as
well as chemical reactors. Particularly important is the way
separation and purification processes underlie our technological
capabilities by enabling us to obtain specific chemical elements and
compounds as needed - another case of adaptive selection (Section
\ref{adapt}), locally going against the grain of the Second Law at
the expense of the environment. The specific processes that enable
these non-unitary
effects are detailed in \cite{HenSeaRop11}.\\

 Molecular and solid state systems inherit the arrow of time from their
constituents (current flows, rectifiers, gates), leading to
electrical and electronic systems \cite{Hug08}, computers
\cite{Tan90}, and nanotechnology devices \cite{Zim79}. The flow of
time in the response of the components at each level is used by the
designers in making time-responsive higher circuit elements; they
are all classical elements, so at their operation emerges out of
state-vector reduction at the micro level.

\subsubsection{Geological and astrophysical arrow of time} Diffusion
plays a crucial role in the environment, where it relates to
pollution hazards in the ocean, atmosphere, rivers, and lakes
\cite{Csa73}.\\

It occurs in astrophysics, where the Fokker-Planck equation implies
continuous production of entropy (\cite{Sas87}: Chapter 4). One-way
processes based in the underlying radiative processes and diffusion
are important in stars, galaxies, radio sources, QSOs, and so on
\cite{Ree95,Shk78}. In particular supernovae are irreversible
processes:  gravity creates order (attractor of dynamic system:
direction comes from initial conditions, SN explosion creates
disordered state irreversibly. This is crucial to the formation of
planets round second generation stars that can become homes for life
(\cite{Sil01}: 345-350).

\subsubsection{Biological arrow of time} Biology takes the arrow of
time in lower level processes for granted. These time asymmetric
processes, e.g. cell processes, synaptic process, propagate their
arrow of time up the hierarchy \cite{CamRee05} to the macro states.
Thus diffusion across synaptic cleft, propagation of action
potential from dendrite to axons, the rectifying action of
voltage-activated channels underlying the nerve impulse, lead to
time asymmetric brain processes. Micro asymmetry in these processes
results in emergent time asymmetry in macro events in the brain:
emergent structures in biology inherit their arrow of time from the
underlying modules and
their non-equilibrium interactions.\\

At the community level, crop ecology depends on the one-way process
of leaf photosynthesis (\cite{LooCon92}:257-288) and its consequences
for plant growth. Dissipative forces in the interaction of
components must be modeled in looking at energy flows in ecosystems
\cite{Odu72} and in physical processes such as weathering, erosion,
and deposition (\cite{WhiMotHar84}:225-249,274-280). A
thermodynamically based arrow of time is involved, starting at the level of grains of sand
and dust particles, and cascading up to weathering of mountains and global spread of pollution.\\

As for evolution, there is a time asymmetry in the Darwinian
evolutionary process: initial conditions on earth were very simple
in biological terms, so there was no way but up!  At a macro level,
fitness flux characterizes the process of natural selection
\cite{MusLas10} and satisfies a theorem that shows existence of a
fitness flux even in a non-equilibrium stationary state. As always,
adaptive selection is a non-unitary process. In such processes,
energy rate density serves as a plausible measure of complexity
\cite{Cha98,Cha10}, but the process is not necessarily always up:
evolution is an imperfect ratchet.

\subsubsection{Experience and Memory} The process of experiencing and
remembering is based in underlying time-asymmetric synaptic
processes \cite{KanSchJes00} that lead to the experience of
subjective time \cite{Gra11}.\\

\textbf{Overall}, lower level entities experience an arrow of time
because of their context; as they act together to create higher
level entities, they inevitably export this arrow of time into the
higher level structures that emerge.

\section{The nature of spacetime} \label{sec:time1}
The passage of time is crucial to the understanding of physical
reality presented here: not just as a subjective phenomenon related
to the mind, but as an objective phenomenon related to physical
processes occurring from the very early universe to the present day.
The best spacetime model for what occurs is an evolving block
universe (Section \ref{EBU}), increasing with time from the start of
time until the end of time. This provides the ultimate source of the
microphysics arrow of time (section \ref{sec:arrow}), as well as a
solid reason for preserving causality by preventing existence of
closed timelike lines (Section \ref{sec:closed}).

\subsection{The Evolving Block Universe}\label{EBU} The passage of time is a
real physical processes, as exemplified in all the cases
discussed above.  Our spacetime picture should adequately reflect this fact.\\

\textbf{The nature of existence is different in the
past and in the future } - Becoming has meaning
(\cite{Edd28}:94-110). Different ontologies apply in the past and
future, as well as different epistemologies.\\

 \noindent One can
express this essential feature by viewing spacetime as an
\emph{Evolving Block Universe} (EBU) \cite{Taoo00,Ell08}. In such a
view the present is different from the past and the future; this is
represented by an emergent spacetime which grows with time, the
present separating the past (which exists) from the future, which
does not yet exist and so does not have the same ontological status.
The past is the set of events that have happened and so are
determined and definite; the future is a set of possibilities that
have not yet happened. The present separates them, and the passage
of time is the continual progression by which the indeterminate
becomes determinate. This is not derived from the physics equations, but
postulated independently as the way the function.\\

It must be emphasized here that it is not just
the contents of spacetime that are determined as time evolves; the
spacetime structure itself also is only definite once events have
taken place. For example quantum fluctuations determined the
spacetime inhomogeneities at the end of inflation \cite{Ell08};
hence they were intrinsically unpredictable; the outcome was
only determined as it happened. The part of spacetime that \emph{exists} at any instant is the past
part of spacetime, which continually grows. This is the evolving
block universe. The future is a possibility space, waiting to be
realized. It does not yet exist, although it is not generic: there
are a restricted set of possibilities that can emerge from any
specific present day state. Classically they would be determined,
but irreducible quantum uncertainty prevents unique predictions
(Section \ref{sec:basic}; \cite{Ell11a}). This is where the
essential difference between the future and the past arises.\\

Thus in Figure 7, the time up to the present has happened, and
everything to the past of the present is determined. The time to the
future of the present has yet to occur; what will happen there is
not yet determined. The present time is a unique aspect of spacetime
at each instant, forming the future boundary of spacetime, and it
keeps moving up as time elapses. It is determined by integrating
proper time along the fundamental world lines defined by being Ricci
eigenlines \cite{Ell13}, because causation takes place along
timelike lines, not spacelike surfaces; this gives the usual
surfaces of constant time in a Robertson-Walker spacetime. The
relativity of simultaneity is a psychological construct that is
irrelevant to physical processes,
and so that issue has no physical import \cite{Ell13}. \\

-----------------------------------------------------------------------------------\\
\begin{figure}
\begin{center}
\label{FIGURE 7}
 \includegraphics[width=7.0in]{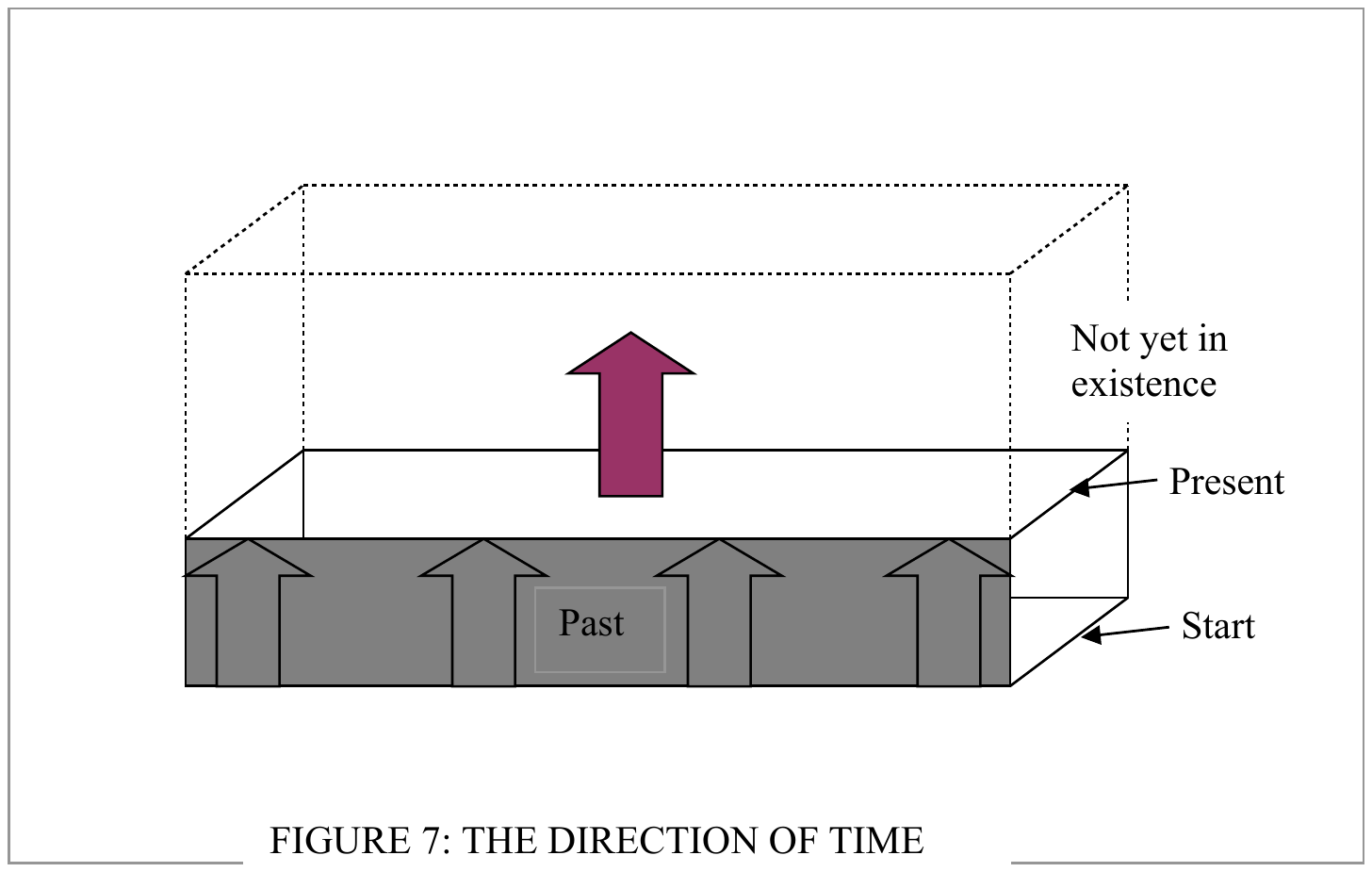}
\end{center}
\end{figure}

FIGURE 7 HERE\\

Caption:  THE DIRECTION OF TIME: \\

\texttt{{\small The direction of time results from the fact that at any specific time $t_0$,
spacetime exists in the past ($t< t_0$) but not in the future ($t> t_0$). The direction of time --
the direction in which spacetime is growing -- points for the start of the universe to the
growing boundary at the present.}}\\

-----------------------------------------------------------------------------------\\

It is the viewpoint of this paper that at a micro level, time passes
as effective wave function collapse takes place: the indefinite
future changes to the determined past as this happens (equation
(\ref{trans})). Alternative views abound, but this is the view I
propose here; whatever underlying theory one may have, this is the
effective theory that must emerge. This is possibly best represented
as a \emph{Crystalizing Block Universe} (CBU) when we take quantum
effects such as entanglement in time into account \cite{EllRot10}.
When one coarse grains the local micro time determined in this way,
related to physical processes by (\ref{evolution}),  it will lead to
the macro time parameter.

\subsubsection{The preferred world lines}
The past/future cut continually changes with time, at any specific
time defining the present, so it is fundamental to physical
processes;  how is it determined? Physical processes are based in
timelike and null worldlines rather than spacelike surfaces, so this
process of becoming determinate happens along preferred timelike
worldlines in spacetime, as a function of proper times along those
world lines since the beginning of the universe, augmented by
processes along null geodesics(in the real universe, timelike
effects dominate at most places at both early and late times). The
metric function determines proper time for an observer along each
observer's world line, as a line integral along their timelike world
lines - this is a basic feature of general relativity
\cite{HawEll73}.  This provides the proper time parameter $\tau$ at
each event that determines the rate at which physical processes
happen through local dynamical equations such as the Schr\"{o}dinger
equation (\ref{evolution}) at the micro level, and the Maxwell
equations and Einstein field equations in the 1+3
covariant form \cite{EllMaaMac11} at the macro level .\\

The proposal is that on a large scale, what matters is the average
motion of all matter present in an averaging volume, which
determines the average 4-velocity of matter in the universe
\cite{Ell71}, so this is what determines the communal cosmological
time. The surfaces of constant time $S(\tau)$ will be determined by
the integral of proper time $\tau$ along the timelike eigenvectors
of the total matter stress tensor $T_{ab}$ from the start of the
universe to the present time \cite{Ell13}. Through the Einstein
equations \cite{HawEll73}, these curves, representing the average
motion of matter at each event determined on a suitable averaging
scale, will be the timelike eigencurves of the Ricci tensor
$R_{ab}$, which will be uniquely defined in any realistic spacetime
(the real universe is not a de Sitter or Anti de Sitter
spacetime).\footnote{There are no preferred worldlines or space
sections in special relativity, so the idea does not work in that
context; this has been used as an argument against the EBU idea. But
it is general relativity that determines the spacetime in the real
universe via the Einstein Field Equations, so this special
relativity indeterminacy is not the
physically relevant case and the argument does not apply.} \\

This construction non-locally defines a unique surface $S(\tau_0)$ (`the present') where
the transition event is taking place at any specific time $\tau_0$; spacetime is
defined for $0 < \tau < \tau_0$, but not for $\tau > \tau_0$.
These surfaces are derivative rather than fundamental: as indicated,
the essential physical processes take place along timelike world
lines. These surfaces of transition need not be instantaneous for
the preferred world lines, and are not even necessarily spacelike.
Their existence breaks both Lorentz invariance and general
covariance; that is fine, this is an unsurprising  case of a broken
symmetry, as occurs in any specific realistic solution of the field
equations (for example any perturbed Friedmann-Lema\^{\i}tre
spacetime \cite{Dod03,EllMaaMac11}). The existence of physical objects
is related to conservation relations between entities at
successive times, entailing continuity of existence between
correlated sets of properties as time progresses.\\

The quantum measurement process (i.e. effective projection of the
wave function to an eigenstate with specific values for the relevant
variables) is associated with specific local physical entities such
as a particle detector or photon detector or rhodopsin molecules
\cite{Ell11a}, for these determine what happens. Thus we may expect
that the relevant world lines for the quantum to classical
transition (superpositions changing to eigenstates) will be fixed by
the local motion of matter: on a small scale, that of a detection
apparatus; on a larger scale, the average motion of matter in a
large averaging volume.

\subsubsection{Proposal}
In summary, at any instant the ontological nature of the past,
 present, and future is fundamentally different.This determines the
 direction in which time flows.
\begin{itemize}
\item  \textbf{Viewpoint: The evolving nature of space time}: the proposal is that spacetime is
an \emph{Evolving Block Universe},
  where the essential difference between the past (it exists) and future (it does not yet exist) generates
 a time asymmetry in all local physical processes and so creates the direction of time (Figure 7).
 Spacetime starts at the beginning of the universe and then grows steadily
 until the end of time; this direction then cascades down to determine the
 arrow of time in local systems (Figure 3). 
 \end{itemize}
It has been suggested to me that this EBU proposal is a
philosophical position. I disagree: it is a selection principle for
viable cosmologically relevant space times, in much the same way
that one insists that the speed of sound $v_s$ must be less than the
speed of light $c$ in any viable solution. It has mathematical
outcomes, as explained above: at any specific time $\tau_0$,
spacetime is defined for $0 < \tau \leq \tau_0$, but not for $\tau >
\tau_0$; hence any integrals for fields or radiation on the surface
$S(\tau_0)$ can only range over values $\tau < \tau_0$. This
excludes advanced solutions of the wave equation for any variable,
and only the retarded Feynman propagator \cite{FeyHib65} will make
physical sense, because you can't integrate over the future domain
if it does not yet exist. \\

The way  this time asymmetry ``reaches down'' to the quantum
measurement process and the state preparation process is still to be
clarified. The working hypothesis is that it must do so, determining
the local quantum arrow of time locally in each domain in such a way
that they do indeed add up to a coherent global arrow of time. This
is clearly a speculation, but it sets a possible agenda for
investigation.

\subsubsection{The contrary view} This view is of course contrary to that expressed by some
philosophers (e.g. \cite{Pri96}) and by many quantum physicists,
particularly related to the idea of the wave function of the
universe and the Wheeler-de Witt equation (e.g. Barbour
\cite{Bar99}). In \cite{Ell11a}, I claimed that the basis for
believing in that approach is not on a solid footing. In brief, the
argument is, \emph{We have no evidence that the universe as a whole
behaves as a Hamiltonian system}. Indeed, because the behavior of
the universe as a whole emerges from the conjunction in complex
configurations of the behavior of its components, it is likely that
this is not true, except perhaps at the very earliest times before
complex configurations existed \cite{Ell11a}.\\

This counter viewpoint is put in many articles and books, stating that
every event in the past and future is implicit in the current
moment, because that is what the equations say; either time does not exist, or it does
not flow \cite{Dav12,Pri96,Bar99}.\\

But the question is which equations, and when are they applicable, and what is their context of application?
As emphasized so well by Eddington \cite{Edd28}:246-260), our
mathematical equations representing  the behavior of macro objects
are highly abstracted version of reality, leaving almost all the
complexities out. The case made in \cite{Ell11a} is that when true
complexity is taken into account, the unitary equations leading to
the view that time is an illusion are generically not applicable
except to isolated micro components of the whole; \cite{Ell13} shows an alternative coordinate system where the Hamiltonian does not vanish. The counter viewpoint
expressed often supposes a determinism of the future that is
not realized in practice, denying the
applicability of quantum uncertainty to the real universe. But that
uncertainty is a well-established fact \cite{FeyLeiSan65,GreZaj06},
which can have macroscopic consequences in the macro world, as is
demonstrated by the historic process of structure formation
resulting from quantum fluctuations during the inflationary era
\cite{Dod03}. These inhomogeneities were not determined until the
relevant quantum fluctuations had occurred, and then become
crystalized in classical fluctuations; and they were unpredictable,
even in principle.

\subsection{Closed timelike lines} \label{sec:closed}

A longstanding problem for general relativity theory is that closed
timelike lines can occur in exact solutions of the Einstein Field
Equations with reasonable matter content, as shown famously in the
static rotating G\"{o}del solution \cite{HawEll73}. This opens up
the possibility of many paradoxes, such as  killing your own
grandparents before you
were born and so creating causally untenable situations. \\

It has been hypothesized that a \emph{Chronology Protection
Conjecture} \cite{Haw92} would prevent this happening. Various
arguments have been given in its support \cite{Vis02}, but this
remains an \emph{ad hoc} condition added on as an extra requirement
on solutions of the field equations, which do not by
themselves give the needed protection.\\

The EBU automatically provides such protection \cite{Ell13}, because creating
closed timelike lines in this context requires the determined part
of spacetime intruding on regions that have already been fixed. But
the evolving spacetime regions can never intrude into the completed
past domains and so create closed timelike lines, because to do so
would require the fundamental world lines to intersect each other;
and that would create a space-time singularity, because they are the
timelike eigenvectors of the Ricci tensor, and in the real universe,
there is always matter or radiation present. The extension of time
cannot be continued beyond such singularities, because they are the
boundary of spacetime \cite{Ell11}.

\begin{quote}
 \textbf{Causality}: \emph{The existence of closed timelike lines
(\cite{Car10}:93-116) is prevented, because if the fundamental
world lines intersect, a spacetime singularity occurs
\cite{HawEll73}: the worldlines are incomplete in the future, time
comes to an end there, and no ``Grandfather Paradox'' can occur}.
\end{quote}

\section{The Arrow of Time}\label{sec:arrow}

In an evolving block universe, where the flow of time is real, one cannot resolve the arrow of time problem through the idea \textbf{AT1}: there are different conditions in the far future and
the far past. It does not apply, because that cannot be applied if the
future does not yet exist. The solution is rather the combination of
\textbf{AT3}, setting the master direction of time at the
cosmological scale, in combination with the speciality condition \textbf{AT3}, which
 validates the second law of thermodynamics. It propagates down to give an arrow of time at each lower level by setting special environmental contexts at each level,  and then propagates up in emergent structures, to give effective
time asymmetric laws at each level. \\

Together these create the EBU where the arrow of time is built in to
the fact that the past has taken place,  and the future is yet to
come; the past exists as what has happened, the future as
(restricted) potentialities.
\begin{itemize}
\item Only radiation from the past can affect us now, as only the past has happened.
Radiative energy arrives here and now
from the past null cone, not the future null cone (Figure 3).
\item Only the retarded Feynman Green's function makes physical
sense, because only the past can send causal influences to us. This
solves the issue of the local electromagnetic arrow of time.
\item The matter that exists here and now was created by nucleosynthesis in the past (Figure 4).
It bears in its very existence a record of the events of
cosmological and stellar nucleosynthesis.  Future potentialities are unable to influence us in this way.
\item We can influence the future by changing conditions as to what will happen then;
we cannot do so for the past, as it has already occurred. The  relevant wave function
has already collapsed and delivered a specific result.
\item Overall, the micro laws of physics are time symmetric, for example Feynman
diagrams can work in both directions in time, but the context in which they operate
(the EBU) is not.  Thus their outcome of necessity has a determinate arrow of time,
which underlies the very concept of causation as we know it. If this was not so,
cause and effect would not be distinguishable.
\end{itemize}

\subsection{The top-down and bottom up cascades}\label{time10}
The overall picture that emerges is shown in Diagram 2.\\
\begin{center}
\begin{tabular}{|c|c|c|}\hline
   \multicolumn{3}{|c|} {\emph{\textbf{The Arrow of Time}} }
   \\ \hline
   \textbf{Cosmology} &   &  \textbf{Brain, Society} \\ \hline
   \emph{Top-down effects }  $\Downarrow$ &      & $\Uparrow$ \emph{Bottom-up effects }   \\ \hline
   \textbf{Non-equilibrium environment} &  $\Rightarrow$  & \textbf{Molecular processes} \\ \hline
    \emph{Top-down effects } $\Downarrow$ & & $\Uparrow$ \emph{Bottom-up effects } \\ \hline
   \textbf{Quantum Theory} &  $\Rightarrow$  & \textbf{Quantum Theory} \\ \hline
\end{tabular}\\
\end{center}
{}\\ \textbf{Diagram 2:} \emph{Contextual determination of the
arrow of time cascades down from cosmology to the underlying micro
processes, on the natural sciences side,
and then up to the brain and society, on the human sciences side}.\\

In summary:
\begin{itemize}
\item Spacetime is an evolving block
  universe, which grows as time evolves. This fundamental arrow of
  time was set at the start of the universe.
  \item The observable part of the universe started off in a special state
  which allowed structure formation to take place and entropy to
  grow.
  \item The arrow of time cascades down from cosmology to the quantum
level (top down effects) and then cascades up in biological systems
(emergence effects), overall enabled by the expanding universe
context leading to a dark night sky allowing local non-equilibrium
processes to occur.
\item There are an array of
technological and biological mechanisms that can detect the
direction of time, measure time at various levels of precision, and
record the passage of time in physically embodied memories.
  \item These are irreversible processes that occur at the classical level,
  even when they have a quantum origin such as a tunneling process, and so at a foundational level must
be based in the time-irreversible quantum measurement process.
  \item  In conceptual terms they are the way the arrow of time parameter $t$ in the basic
equations of physics (the Dirac and Schr\"{o}edinger equations
(\ref{evolution}), Maxwell's equations and Einstein's equations on
the 1+3 covariant formulation \cite{EllMaaMac11}) is realized and
determines the rate of physical processes and hence the way time
emerges in relation to physical objects.
  \item Each of these processes is enabled by top-down action taking
  place in suitable emergent local structural contexts, provided by
  molecular or solid-state structures. These effects could not occur in
  a purely bottom-up way.
\end{itemize}

\subsection{A contextual view of the arrow of time}\label{sec:conclusion}
This paper has extended the broad framework of
\cite{Ell11a} to look in detail at the issue of the arrow of time.
It has made the case that this is best looked at in terms of the
hierarchy of complexity (Table 1), where both bottom-up and top-down
causation occur. Detailed examples have been given of how this works
out in terms of arrow of time detectors, clocks, and records of past
events. \textbf{AT3} sets  the master arrow of time. The EBU starts at the beginning of time, the future direction of time is that direction in which spacetime is growing.
\begin{quote}
\textbf{The Arrow of Time}:\emph{ On the view presented here, the ultimate
resolution of the Arrow of Time issue is provided by the fact we
live in an evolving block universe starting from an initial singularity. Only the past can
influence us, because the future does not yet exist, so it cannot
causally affect us.}
\end{quote}
It then cascades down in physical systems, allowing entropy to grow because of the past condition \textbf{AT2}, and the up in biological systems, allowing complexity to emerge because adaptive selection takes place. But these are not the basic source of the arrows at each level of the hierarchy: they are effects of the fundamental cause.\\

Key to this is the time-asymmetry of the quantum measurement
process, which I suggest emerges in a contextual way.

Firstly, a detection process depends on setting the detector into a ground state
before detection takes place (analogously to the way computer memories have to be
notionally cleared before a calculation can begin). This is an asymmetric adaptive
selection process, because what is needed is kept and what is not needed is discarded,
whereby any possible initial state of the detector is reduced to a starting state,
thereby decreasing entropy. It will be implemented as part of the detector design.

Secondly, one might suggest that the asymmetry of the collapse may
derive from the fact that the future does not yet exist in a EBU
(Section \ref{sec:time1}), and this is the time-asymmetric context in
which local any physical apparatus or other context leads to a constrained set of outcomes
by their specific construction. There does not
seem to be any other plausible way to relate the global cosmological
arrow of time to the local arrow of time involved in collapse of the
wave function. How this happens needs to be elucidated, as part of
the investigation of how state vector collapse takes place as a
contextually dependent process in specific physical contexts \cite{Ell11a}. \\

\noindent \textbf{Acknowledgements}: \\

I thank Max Tegmark and Anthony Aguirre for organizing a very useful
meeting of the FQXI Institute on The Nature of Time, the
participants at that meeting for many interesting presentations, and
Paul Davies for very helpful discussions. I thank the National
Research Foundation (South Africa) and the University of Cape Town
for support, and the referee for a careful reading of the paper that
has led to significant improvement.

\newpage

gfre::version 2013-03-01
%

\begin{thebibliography}{99}

\bibitem{AhaRoh05} Y Aharaonov and D Rohrlich (2005) \emph{Quantum paradoxes} (Weinheim: Wiley-VCH).

\bibitem{Alb00} D Albert (2000). \emph{Time and Chance} (Cambridge, MA: Harvard University
Press).


\bibitem{AloFin71} M Alonso and E J Finn (1971) \emph{Fundamental University
Phyiscs III: Quantum and Statistical Physics}
 (Reading, Mass: Addison Wesley).

\bibitem{Atk94} P W Atkins (1994) \emph{Physical Chemistry} (Oxford: Oxford
University Press).

\bibitem{And72} P W Anderson(1972) ``More is Different'' \emph{Science} \textbf{177}, 377.
Reprinted in \emph{P W Anderson: A Career in Theoretical Physics}.
(World Scientific, Singapore. 1994).

\bibitem{AulEllJae08} G Auletta, G
Ellis, and L Jaeger (2008) ``Top-Down Causation: From a
Philosophical Problem to a Scientific Research Program'' \emph{J R
Soc Interface} \textbf{5}: 1159-1172
[http://arXiv.org/abs/0710.4235].

\bibitem{Bar99} J B Barbour (1999) \emph{The End of Time: The Next Revolution
in Physics} (Oxford: Oxford University Press).

\bibitem{Ben03} C H Bennett (2003) ``Notes on Landauer's principle, Reversible Computation and Maxwell's Demon''. \emph{Studies in History and Philosophy of Modern Physics} \textbf{34}: 501–510.

\bibitem{BrePet06} H.-P Breuer and F Petruccione (2006) \emph{The Theory of open quantum
systems} (Oxford: Clarendon Press).

\bibitem{Buz06}
G Buzsaki (2006) \emph{Rhythms of the Brain} (Oxford: Oxford
University Press).

\bibitem{Cal11} C Callender (2011), ``Thermodynamic Asymmetry in Time'', \emph{The Stanford
Encyclopedia of Philosophy} (Fall 2011 Edition), Edward N. Zalta
(ed.), http://plato.stanford.edu/entries/time-thermo/.

\bibitem{CamRee05} N  A Campbell and J B Reece (2005) \emph{Biology}
(Benjamin Cummings).

\bibitem{Car10} S Carroll (2010) \emph{From Eternity to here: the quest for the ultimate arrow of time} (New York: Dutton).

\bibitem{CarTam10}  S  M Carroll and  H Tam (2010) ``Unitary Evolution and Cosmological Fine-Tuning''
 arXiv:1007.1417.

\bibitem{Cha98} E J Chaisson (1998) ``The cosmic environment for the growth of
complexity'' \emph{Biosystems} \textbf{46}:13-19.

\bibitem{Cha10} E Chaisson (2010) ``Energy Rate
Density as a Complexity Metric and Evolutionary Driver''
\emph{Complexity} \textbf{16}: (3)


\bibitem{Csa73} G T Csanady (1973) \emph{Turbulent Diffusion in the
Environment} (Dordrecht: D Reidel).

\bibitem{Dav74} P C W Davies (1974) \emph{The Physics of Time Asymmetry}
(Surrey University Press).

\bibitem{Dav12} P C W Davies (2012) ``That Mysterious Flow''. \emph{Scientific
American} Special Edition: \emph{A Matter of Time} Vol \textbf{21}:
2012), 8-13.


\bibitem{Dod03} S Dodelson (2003) \emph{Modern Cosmology} (New York: Academic Press).

\bibitem{Dur00} A Durrant (2000) \emph{Quantum Phyiscs of Matter}
(Bristol: Institute of Physics and The Open University).

\bibitem{Edd28}
A S Eddington (1928). \emph{The Nature of the Physical World}. (London: MacMillan)

\bibitem{Ell71} G F R Ellis (1971) ``Relativistic Cosmology". In
\emph{General Relativity and Cosmology},
Ed. R K Sachs (Academic Press, 1971), 104-179.
Reprinted 
\emph{Gen. Rel. Grav}. \textbf{41}: 581 (2009).

\bibitem{Ell84} G F R Ellis (1984) ``Relativistic cosmology: its nature, aims and problems''.
In \emph{General Relativity and Gravitation}, Ed B Bertotti et al (Reidel), 215-288.

\bibitem{Ell95}
G F R Ellis (1995): ``Comment on `Entropy and the Second Law: A
Pedagogical alternative', By Ralph Baierlein''. \emph{Am Journ Phys}
\textbf{63}:472.


\bibitem{Ell04} G F R Ellis (2004): ``True Complexity and its Associated Ontology''.
In \emph{Science and Ultimate Reality}. Ed. J D Barrow, P C W
Davies, and C L Harper (Cambridge: Cambridge University Press),
607-636.

\bibitem{Ell06} G F R Ellis (2006) ``Physics in the Real Universe: Time and
Spacetime''. \emph{GRG } \textbf{38}:1797-1824
[http://arxiv.org/abs/gr-qc/0605049].

\bibitem{Ell06a} G F R Ellis (2006) ``Issue in the Philosophy of cosmology'' In
\emph{Handbook in Philosophy of Physics}, Ed J Butterfield and J
Earman (Elsevier, 2006), 1183-1285
[http://arxiv.org/abs/astro-ph/0602280].

\bibitem{Ell08} G  F R Ellis (2008) ``On the nature of causation in complex
systems'' \emph{Trans Roy Soc South Africa} \textbf{63}:
69-84.

\bibitem{Ell11} G  F R Ellis (2012) ``Top down causation and emergence: some comments on
mechanisms'' \emph{Journ Roy Soc Interface} (London) \textbf{2}:
126-140.

\bibitem{Ell11a} G  F R Ellis (2011a)
``On the limits of quantum theory: contextuality and the
quantum-classical cut'' [arXiv:1108.5261].

\bibitem{Ell13} G F R Ellis (2013) ``Space time and the passage of time''
For \emph{Springer Handbook of Spacetime} ed V Petkov (Heidelberg:
Spriniger) [arXiv:1208.2611].

\bibitem{EllNobOCo11} G F R Ellis, D Noble, and T
O'Connor (2011) (Eds) Special issue \emph{Journ Roy Soc Interface
Focus} (London) on top down causation, to appear.

\bibitem{EllBru89} G F R Ellis and M Bruni (1989) ``A covariant and gauge-free approach to
density fluctuations in cosmology'' \emph{Phys Rev} \textbf{D40}:
1804-1818.

\bibitem{EllMaa04} G F R Ellis and R Maartens (2004)
``The Emergent Universe: inflationary cosmology with no
singularity'' Class. Quant. Grav. 21: 223-232 [gr-qc/0211082].

\bibitem{EllMaaMac11} G F R Ells, R Maartens
and M A H MacCallum (2011) \emph{Relativistic Cosmology} (Cambridge:
Cambridge University Press).

\bibitem{EllMurTsa04} G F R Ellis, J Murugan, and C G Tsagas (2004) ``The Emergent Universe:
An Explicit Construction'' \emph{Class. Quant. Grav}. \textbf{21}:
233-249' [gr-qc/0307112].

\bibitem{EllRot10} G F R Ellis and T Rothman (2010) ``Crystallizing block universes''.
\emph{International Journal of Theoretical Physics} \textbf{49}: 988
[http://arxiv.org/abs/0912.0808].

\bibitem{EllSci72} G F R Ellis and D W Sciama (1972) ``Global and non-global problems in
cosmology''. In \emph{General Relativity (A Synge Festschrift)}, ed.
L. O'Raifeartaigh (Oxford: Oxford University Press), 35-59.

\bibitem{EllSto09} G F R Ellis and W R Stoeger (2009) ``The Evolution of Our Local
Cosmic Domain: Effective Causal Limits'' \emph{Mon Not Roy Ast Soc}
\textbf{398}:1527-1536 [http://arxiv.org/abs/1001.4572].

\bibitem{Fey48}
R P Feynman (1948) ``Space-time approach to non-relativistic quantum mechanics''
\emph{Reviews of Modern Physics} \textbf{20}:367–387.

\bibitem{FeyHib65}
R P Feynman and A R Hibbs (1965) \emph{Quantum Mechanics and Path
Integrals}, Ed D F Styer (Dover: Mineola, New York).

\bibitem{FeyLeiSan63} R P Feynman, R B Leighton and M Sands (1963)
\emph{The Feynman lecturs on Physics: Mainly Mechanics, Radiation,
and Heat} (Reading, Mass: Addison-Wesley).

\bibitem{FeyLeiSan64} R P Feynman, R B Leighton and M Sands (1964)
\emph{The Feynman lecturs on Physics: The Electromagnetic Field}
(Reading, Mass: Addison-Wesley).

\bibitem{FeyLeiSan65} R P Feynman, R B Leighton and M Sands (1965)
\emph{The Feynman lecturs on Physics: Quantum Mechanics} (Reading,
Mass: Addison-Wesley).

\bibitem{Fuc96} H U Fuchs (1996) \emph{The dynamics of Heat} (New York:
Springer).

\bibitem{Gel94}
M Gell-Mann (1994) \emph{The Quark and the Jaguar: Adventures in the
Simple and the Complex} (London: Abacus).

\bibitem{GemMicMah04} J Gemmer, M Michel and G Mahler (2004) \emph{Quantum
Thermodynamics:Emergence of Thermodynamic Behaviour Within Composite
Quantum Systems} (Heidelberg: Springer).

\bibitem{Geo71} N Georgescu-Roegen (1971) \emph{The Entropy Law and the
Ecinomic Process} (Cambridge, Mass: Harvard University Press).

\bibitem{Goeetal02}. P Goettig1, M Groll, J-S Kim, R
Huber and H Brandstetter (2002) ``Structures of the
tricorn-interacting aminopeptidase F1 with different ligands explain
its catalytic mechanism'' \emph{EMBO Journal} \textbf{21}, 5343 -
5352.

\bibitem{Gra11} Gray, P (2011) \emph{Psychology} (New York: Worth).

\bibitem{GreZaj06} G Greenstein and A G Zajonc (2006) \emph{The Quantum Challenge: Modern
Research on the Foundations of Quantum Mechanics} (Sudbury, Mass:
Jones and Bartlett).

\bibitem{Hal03} J Halliwell (2003) ``The interpretation of quantum cosmology and the problem of time''. In \emph{The
Future of Theoretical phyiscs and Cosmology: Celebrating Stephen Hawking's 60th Birthday}, Ed. G W Gibbons,
E P S Shellard and S J Rankin (Cambridge: Cambridge University Press), 675-690.

\bibitem{HalPerZur96} J J Halliwell, J Perez-Mercader, W H Zurek
(Eds) (1996) \emph{Physical Origins of Time Asymmetry} (Cambridge:
Cambridge University Press).

\bibitem{Har87} E R Harrison, \emph{Darkess at night: A Riddle of the
Universe} (Cambridge, Mass: Harvard University Press).

\bibitem{Har00} E R Harrison, \emph{Cosmology: The Science of the Universe}.
(2nd edition) (Cambridge University Press, Cambridge. 2000).

\bibitem{Har03} J Hartle (2003) ``Theories of everything and Hawking's wave function''. In \emph{The
Future of Theoretical phyiscs and Cosmology: Celebrating Stephen Hawking's 60th Birthday}, Ed. G W Gibbons,
E P S Shellard and S J Rankin (Cambridge: Cambridge University Press), 38-49 and 615-620.

\bibitem{HenSeaRop11} E J Henley, J DSeader, and D K Roper (2011)
\emph{Separation Processes and Principles} (Wiley Asia).

\bibitem{Haw92}
S W Hawking (1992) ``The chronology protection conjecture''.
\emph{Phys. Rev}. \textbf{D46}, 603-611.

\bibitem{HawEll73}
S W Hawking and G F R Ellis (1973)  \emph{The Large Scale Structure of Spacetime} (Cambridge: Cambridge BIversity  Press).

\bibitem{Hen06} J Henson (2006) ``The causal set approach to quantum gravity'' In \emph{Approaches to Quantum Gravity - Towards a new understanding of space and time}, Ed. D. Oriti  (Cambridge University Press) [arXiv:gr-qc/0601121].

\bibitem{Hol92} J H Holland (1992) \emph{Adaptaton in natural and artificial systems}
(Cambridge, Mass: MIT Press).

\bibitem{Hug08} E Hughes (2008) \emph{Electrical and Electronic Technology}
(Harlow: Pearson/Prentice Hall).

\bibitem{Ish95} C J Isham (1995) \emph{Lectures on Quantum Theory: Mathematical and Structural
Foundations} (London: Imperial College Press).

\bibitem{ItzZub80} C Itzykson and J-B Zuber (1980) \emph{Quantum Field
Theory} (McGraw Hill).

\bibitem{KanSchJes00} E R Kandel, J H Schwartz, and T M Jessell (2000)
\emph{Principles of Neuroscience} (New York: McGraw Hill).

\bibitem{Kau93} S A Kauffman (1993) \emph{The Origins of Order: Self-Organisation
and Selectionin Evolution} (New York: Oxford).

\bibitem{Kul11}
I M Kulic, M Mani, H Mohrbach, R Thaokar and L Mahadevan (2009)
``Botanical ratchets'' \emph{Proc. R. Soc}. \textbf{B}.

\bibitem{LacMal10}
D Lacoste and K Mallick (2010) ``Fluctuation relations for molecular
motors'' \emph{Biological Physics: Poincare Seminar (2009)}  Ed B
Duplantier and V Rivasseau (Basel: Birkhauser) (61-88)

\bibitem{Lan61} R Landauer (1961) ``Irreversibility and heat generation in the computing process'' \emph{IBM Journal of Research and Development}  \textbf{5}: 183–191.


\bibitem{LefRex90} H S Leff and A F Rex (eds) (1990) \emph{Maxwell's
Demon: Entropy, Information, Computing} (Bristol: Adam Hilger).

\bibitem{Leh73} A L Lehninger (1973) \emph{Bioenergetics} (Menlo Park: W
A Benjamin).

\bibitem{LooCon92} R S Loomis and D J Coonor (1992) \emph{Crop Ecology}
(Cambridge: Cambridge University Press).

\bibitem{Mon73} J L Monteith (1973) \emph{Principles of Environmental
Phyiscs} (London: Edwin Arnold)

\bibitem{Mor90} M A Morrison (1990) \emph{Understanding Quantum Physics: a user's
manual} (Englewood Ciffs: Prentice Hall International).

\bibitem{Muletal05} D J Mulryne, R Tavakol, J E Lidsey, and G F R Ellis
(2005) ``An emergent universe from a loop'' \emph{Phys Rev D} {\bf
71}, 123512 [ astro-ph/0502589].

\bibitem{MusLas10} V Mustonen and M Lässig (2010) ``Fitness flux and ubiquity of
adaptive evolution''  \emph{PNAS} \textbf{107}:4248–4253.

\bibitem{Nicateal01} J G Nicholls, A R Martin, B G Wallace, and P
A Fuchs (2001) \emph{From Neuron to Brain} (Sunderland, Mass:
Sinauer).

\bibitem{Odu72} H T Odum (1972) ``An energy circuit language''. In
\emph{Systems Analysis and Simulation in Ecology}, Vol II, Ed B C
Patten (New York: Academic Press).


\bibitem{Pen89} R Penrose (1989) \emph{The Emperor's New Mind: Concerning Computers, Minds and the
Laws of Physics} (Oxford: Oxford
University Press).

\bibitem{Pen89a} R Penrose (1989) ``Difficulties with inflationary cosmology".
\emph{14th Texas Symposium Relativistic Astrophysics}, ed. E.J.
Fergus (New York Academy of Science, New York)

\bibitem{Pen04} R Penrose (2004) \emph{The Road to Reality: A complete guide to the
Laws of the Universe} (London: Jonathan Cape).

\bibitem{Pen11} R Penrose (2011) \emph{Cycles of Time: An Extraordinary New View of the Universe}
(New York: Knopf).

\bibitem{PesSch95} M E Peskin and D V Schroeder (1995), \emph{An
Introduction to Quantum Field Theory} (Reading, Mass: Perseus
books).

\bibitem{Pri96} H Price (1996) \emph{Time's Arrow and Archimedes' Point} (New York: Oxford
University Press).

\bibitem{Pri07} G N Price, S T Bannerman, E Narevicius, and M G Raizen (2007),
``Single-Photon Atomic Cooling''
\emph{Laser Physics} \textbf{17}:1–4.

\bibitem{Pri08} G N Price, S T Bannerman, K Viering, E Narevicius, and M G Raizen (2008)
``Single-Photon Atomic Cooling'' \emph{Phys Rev Lett}
\textbf{100}:093004.

\bibitem{Rae94} A Rae (1994) \emph{Quantum Physics:
Illusion or Reality?} (Cambridge: Cambridge University Press).

\bibitem{Ree95} M J Rees (1995) \emph{Perspectives in astrophysical
cosmology} (Cambridge: Cambridge University Press).

\bibitem{RhoPfl96} R Rhoades and R Pflanzer (1996) \emph{Human Physiology}
(Fort Worth: Saunders College Publishing).

\bibitem{Rig63} D S Riggs (1963) \emph{The mathematical approach to
physiological problems} (Cambridge, Mass: MIT Press)

\bibitem{Rio98} A. Riotto (1998) ``Theories of Baryogenesis''
[arXiv:hep-ph/9807454].

\bibitem{Roe05} J G Roederer (2005) \emph{Information and its role in Nature} (Heidelberg: Springer).

\bibitem{Roe11} Erik Roeling, W C Germs,B Smalbrugge, E  Geluk, T de Vries, R A J
Janssen  and M Kemerink (2011) ``Organic electronic ratchets doing
work'' \emph{Nature Materials} \textbf{10}: 51–55.

\bibitem{RotEll86} A Rothman and G F R Ellis (1986) ``Can inflation occur in anisotropic
cosmologies?" \emph{Phys Letters} \textbf{B180}:19-24.

\bibitem{RusMugRai06} A Ruschhaupt, J G Muga and M G Raizen  (2006)
``One-photon atomic cooling with an optical Maxwell Demon valve''
\emph{J. Phys. B: At. Mol. Opt. Phys}. \textbf{39}:3833–3838.

\bibitem{Sas87} W C Saslaw (1987) \emph{Gravitational Physics of stellar
and galactic systems} (Cambridge: Cambridge University Press).

\bibitem{Schetal11} G Schaller, C Emary, G Kiesslich, and T  Brandes (2011)
``Probing the power of an electronic Maxwell Demon''
[arXiv:1106.4670v2].

\bibitem{Sch05} E D Schneider and D Sagan (2005) \emph{Into the Cool: Energy
Flow, Thermodynamics, and Life} (Chicago: University of Chicago
Press)

\bibitem{Seretal07}
V Serreli1, C-F Lee, E R Kay, and D A Leigh
(2007) ``A molecular information ratchet'' \emph{Nature}
\textbf{445}: 523-527

\bibitem{Shk78} I S Shklovskii (1978) \emph{Stars: Their Birth, Death, and
Life} San Francisco: Freeman)

\bibitem{Sil01} J Silk (2001) \emph{The Big Bang} (New York:
Freeman).

\bibitem{Tan90}  A S Tanenbaum (1990) \emph{Structured Computer Organisation} (Englewood Cliffs: Prentice Hall).

\bibitem{Taoo00} M Tooley (2000) \emph{Time, Tense, and Causation} (Oxford: Oxford University Press)

\bibitem{Truetal95} M C Trudeau, J W Warmke, B Ganetzky, and G A Robertson (1995) ``HERG, a human inward rectifier in the voltage-gated potassium channel family'' \emph{Science} \textbf{269}: 92–5.

\bibitem{TurSch79} M S Turner and D Schramm (1979) ``Cosmology and Elementary Particle
Phyiscs'' \emph{Physics Today} (September 1979). Reprinted in D N
Schramm, \emph{The Big Bang and other explosions in nuclear and
particle astrophysics} (Singapore, World Scientific: 1996).


\bibitem{Vis02}
M Visser (2002) ``The quantum physics of chronology protection''. In
\emph{The Future of Theoretical Physics and Cosmology: Celebrating
Stephen Hawking's 60th Birthday} Ed G W Gibbons, E P S Shellard and
S J Rankin (Cambridge: Cambridge University Press), 161-173
[arXiv:gr-qc/0204022v2].


\bibitem{Wal01}
D  Wallace (2002) ``Worlds in the Everett Interpretation''
\emph{Studies in the History and Philosophy of Modern Physics}
\textbf{33}:637--661.

\bibitem{Wat70} J D Watson (1970) \emph{The Molelcular Biology of the Gene} (Menlo Park: W A Benamin).

\bibitem{Whe78} J A Wheeler (1978), ``The 'Past' and the 'Delayed-Choice
Double-Slit Experiment',''  in A.R. Marlow, editor,
\emph{Mathematical Foundations of Quantum Theory} (New York:
Academic Press), 9–48.

\bibitem{WheFey45} J A Wheeler and R P Feynman (1945), ``Interaction with the
Absorber as the Mechanism of Radiation''. \emph{Rev. Mod. Phys}.
\textbf{17}: 157-181 .

\bibitem{WhiMotHar84} I D White, D N Mottershead, and S J Harrison
(1984): \emph{Environmental systems} (London: Unwin Hyman)

\bibitem{Wei95} S Weinberg (1995) \emph{The Quantum Theory of Fields Volume I} (Cambridge: Canbridge University Press).

\bibitem{Wik11}
Wikibooks: \emph{Energy in Ecology}, Chapter 14.

\bibitem{WisMil10}
H M Wiseman and G J Milburn (2010) \emph{Quantum Measurement and Control}
(Cambridge: Cambridge University Press).

\bibitem{Woo05} B Wood (2005) \emph{Human Evolution: A Very Short
Introduction} (Oxford: Oxford University Press).

\bibitem{Zee03} A Zee (2003) \emph{Quantum Field Theory in a nutshell}
(Princeton: Princeton University Press).

\bibitem{Zeh07} H-D Zeh (2007) \emph{The Physical Basis of the Direction of Time} (Berlin: Springer
Verlag).

\bibitem{Zeh90} H D Zeh (1990) ``Quantum measurement and entropy''
In \emph{Complexity, Entropy and the Physics of Information}, Ed W H
Zurek (Redwood City: Addison Wesley), 405-421.

\bibitem{Zim79} J M Ziman (1979) \emph{Principles of the theory of solids}
(Cambridge: Cambridge University Press).

\end{thebibliography}
\end{document}